\documentclass[times,twocolumn,final]{elsarticle}

\usepackage{framed,multirow}

\usepackage{amssymb}
\usepackage{latexsym}
\usepackage{comment}
\usepackage{amsmath}
\usepackage{amsfonts}
\usepackage{booktabs}
\usepackage{algorithm}
\usepackage{algorithmic}
\usepackage{color}
\usepackage{mathtools}
\usepackage{mathrsfs}
\usepackage{subfigure}

\usepackage{url}
\usepackage{xcolor}
\definecolor{newcolor}{rgb}{.8,.349,.1}

\usepackage{hyperref}

\usepackage[switch,pagewise]{lineno} 

\journal{Computers \& Graphics}

\begin{document}


\begin{frontmatter}
 
\title{GAN-based Reactive Motion Synthesis with Class-aware Discriminators for Human-human Interaction}
\author[1]{Qianhui Men\corref{cor1}}
\cortext[cor1]{Corresponding author: Qianhui Men}
\emailauthor{qianhumen2-c@my.cityu.edu.hk}{Qianhui Men}
    
\author[2]{Hubert P. H. Shum} 
\emailauthor{hubert.shum@durham.ac.uk}{Hubert P. H. Shum}

\author[3]{Edmond S. L. Ho}
\emailauthor{e.ho@northumbria.ac.uk}{Edmond S. L. Ho}

\author[1]{Howard Leung}  
\emailauthor{howard@cityu.edu.hk}{Howard Leung}

\address[1]{Department of Computer Science, City University of Hong Kong, Hong Kong, China}
\address[2]{Department of Computer Science, Durham University, Durham, United Kingdom}
\address[3]{Department of Computer and Information Sciences, Northumbria University, Newcastle upon Tyne, United Kingdom}


\begin{abstract}

Creating realistic characters that can react to the users' or another character's movement can benefit computer graphics, games and virtual reality hugely. However, synthesizing such reactive motions in human-human interactions is a challenging task due to the many different ways two humans can interact. While there are a number of successful researches in adapting the generative adversarial network (GAN) in synthesizing single human actions, there are very few on modelling human-human interactions. In this paper, we propose a semi-supervised GAN system that synthesizes the reactive motion of a character given the active motion from another character. Our key insights are two-fold. First, to effectively encode the complicated spatial-temporal information of a human motion, we empower the generator with a part-based long short-term memory (LSTM) module, such that the temporal movement of different limbs can be effectively modelled. We further include an attention module such that the temporal significance of the interaction can be learned, which enhances the temporal alignment of the active-reactive motion pair. Second, as the reactive motion of different types of interactions can be significantly different, we introduce a discriminator that not only tells if the generated movement is realistic or not, but also tells the class label of the interaction. This allows the use of such labels in supervising the training of the generator. We experiment with the SBU and the HHOI datasets. The high quality of the synthetic motion demonstrates the effective design of our generator, and the discriminability of the synthesis also demonstrates the strength of our discriminator. 

\end{abstract}


\end{frontmatter}



\section{Introduction}
\label{sec:1}


Human motion synthesis and generation \cite{shum2008interaction,ho2009character} have benefited the computer animation field. The generation of human reactive motions shows great potentials in controlling the movements of virtual characters in immersive games and human-robot interaction. Given the movement of one character with a 3D pose sequence, reactive motion synthesis aims at generating the movement of the responding character, which responds to the input action. 

While realistic reactive motions can be generated by physical simulation such as ragdoll physics, such an approach is more suitable for creating reactive motions caused by body contact or voluntary movement. On the other hand, human-human interactions cover a wider range of motions that may or may not have any direct contacts. As a result, the kinematic-based approaches \cite{Ho:SIGGRAPH10,ho2013interactive} as well as combined enforcing kinematic and physical constraints \cite{Yin:TVCG2019,Ho:VRST2014} are used for preserving the context in editing close interaction in the literature. Existing work relevant to kinematics-based reactive motion synthesis mainly focus on generating interactions based on the interaction history~\cite{kundu2020cross,baruah2020multimodal}, as well as synthesizing the response with non-parametric algorithms such as Markov Decision Process (MDP)~\cite{huang2014action,shu2016learning,shum12simulating} and motion blending~\cite{komura2005animating,ho2013interactive}. However, it is a challenging task since the reactive motion is expected to respond properly and requires sufficient spatial and temporal synchronizations between the dynamics of the two characters, which can be potentially yet seldom explored by deep learning-based models. 


Deep learning-based models have made motion synthesis task much easier with diverse patterns and styles compounded from large amount of available motion data~\cite{habibie2017recurrent,yu2020structure,battan2021glocalnet}, among which generative adversarial network (GAN) \cite{goodfellow2014generative} has become the most popular~\cite{gui2018adversarial,dong2020adult2child,ferstl2020adversarial} since it is effective in creating vivid samples learned from real distributions. The emergence of conditional GAN \cite{mirza2014conditional} further facilitates the generated samples to meet user's requirements, e.g. generating a specific type of activities \cite{xu2019prediction}, by supervising the generator with the desired label of the generation. While many researches have been found in understanding single human dynamics, adversarial training is less explored in modeling human-human interaction.

In this paper, we propose a semi-supervised GAN system for reactive motion synthesis. The major novelty of the system lies in the purposely designed generator module that model the spatial (i.e. joint movement) and temporal (i.e interaction synchronization) features of the reactive motion, as well as a discriminator that not only tells if a reactive motion is realistic, but also the class label of the interaction. \textcolor{black}{This follows the idea of semi-supervised learning with GAN from~\cite{odena2016semi,kumar2017semi}, where they generate semi-supervised generative framework with an unsupervised discriminator to tell the fidelity of the generation, and a supervised discriminator to tell the class label to enhance the generation with better qualities.}



For the motion generator, we propose an attentive part-based Long Short-Term Memory (LSTM) module, solving the problem to model complicated spatial-temporal correspondence during the interaction. We first propose the spatial structure of the input action by encoding the states of different body parts separately using a hierarchical LSTM layer. Furthermore, we observe that human interaction contains rich spatial and temporal alignments between two characters. When synthesizing interactions, the temporal movements of two characters are prone to be misaligned~\cite{huang2014action,shu2016learning} due to the lack of interactive features modelling. We tackle this problem by constructing an attentive LSTM network in the generator to learn the temporal saliency from the input action, and deliver this time-aware contextual information together with the hierarchical states to help decoding the reaction. The designed temporal attention facilitates the generator to observe the global pattern of input dynamics and perform reactions at the same pace. 


We further propose to embed multi-class classification into the discriminator to endow the generated reactive motion with the property from its interaction type, as inspired by \cite{odena2016semi, kumar2017semi}. This is motivated by the observation that the reactive motion of different class of interaction could be significantly different. In practice, classifying the synthesized reactions increases the capacity of the generator, through generating diverse types of reactive movements. Comparing to conditional GAN that observes the label information in the input stage, our generator can stand alone without prior knowledge of the interaction type while predicting the type-specific reactive dynamics. By sharing partial parameters with a binary classifier, our trained discriminator is capable of improving the reliability of reactive motion given a particular type of incoming motion.


We demonstrate the effectiveness of the proposed reactive motion synthesis method on two popular human-human interaction datasets SBU~\cite{yun2012two} and HHOI~\cite{shu2016learning} which contain many common interaction types such as shaking hands and kicking. The discriminator power is demonstrated by the classification accuracy, and the generator power is demonstrated by the high-quality synthetic motion. 

The main contributions of this research are concluded as follows:
\begin{itemize}
\item We construct a reactive motion synthesis system based on the semi-supervised generative adversarial network.
\item We propose a reactive motion generator with the attentive recurrent network from the part-based body structure to create reactive motion without knowing its interaction category, where the motions of the characters are well-aligned thanks to the attentive module.
\item We propose a dual discriminator with a binary and a multi-class classifier that improves the authenticity and preserves the characteristics of the synthesis from natural reactive behaviors.
\end{itemize}


The rest of the paper is organized as follows: In Section~\ref{sec:related_work}, we review the previous work related to motion representation learning and generation. Section~\ref{sec:preliminaries} and Section~\ref{sec:method} demonstrate the key prior knowledge used in our architecture, and our reactive motion synthesis system, respectively. We further evaluate our synthesized reactive motions and discuss the advantages and limitations in Section~\ref{sec:experiment}. Finally, we make conclusions in Section~\ref{sec:conclusion}.

\section{Related Work}
\label{sec:related_work}
\subsection{Deep Generative Models in Motion Synthesis}
Deep learning-based models are efficient and versatile to generate human movements from vast of motion data. Among deep generative models, motion generation based on Recurrent Neural Network (RNN) becomes the mainstream with its effectiveness in creating sequential movements. With RNN backbones, \cite{battan2021glocalnet} incorporated label information as guidance to synthesize desired future motions based on the initial given poses, and \cite{yu2020structure} retained spatial and temporal structural information in the generated motion using graph convolutional layers. Some researches \cite{habibie2017recurrent,ghorbani2020probabilistic} also adopt variational auto-encoder to learn a competitive motion manifold that can generate stylistic or long-term dynamics with stochastic patterns. Some cutting-edge researches associate deep learning with GAN to predict motion \cite{barsoum2018hp,men2020quadruple} or generate realistic action patterns in videos \cite{yan2017skeleton}. However, they focus on single character synthesis and their generated poses or movements generally contain less variations because of mode collapse.

Some work \cite{kundu2020cross,baruah2020multimodal} adopt RNN to synthesize human-human interaction given the partially observed interaction. \cite{kundu2020cross} synthesized long-term interaction by alternatively generating the pose sequences of the two characters based on the generation history. With such sampling-based manner~\cite{martinez2017human}, errors can be fast accumulated which eventually drifts the generated interaction to a wrong moving direction~\cite{pavllo2018quaternet,pavllo2019modeling}.



\subsection{Spatial Modeling} 
Human action is accomplished by the movements of its articulated joints, and one of the intuitive idea to model the spatial variations of the skeleton joints is to place them in a chain sequence \cite{li2018co}. However, the joints are not physically connected at the margin of each body part, such as foot and head, therefore it may introduce meaningless connection when applying RNN-based sequence learning architecture. To avoid this problem, a graph-based tree structure is proposed \cite{liu2016spatio} to traverse skeleton branches and learn the relationship among adjacent joints. Another solution is to decompose the skeleton structure into valid segments \cite{si2018skeleton,wang21spatiotemporal} to capture low-level limb shifting, and understanding high-level spatial dependencies by concatenating different partitions together.

\subsection{Attention Perception} 
Attention mechanism attends to allocate weights to the valuable content from considerable information, and it shows great advantage especially in context-based sequence learning such as sequence-to-sequence (seq2seq) translation \cite{chiu2018state}. The translated sample can be aligned as the focus of the decoder will be updated during the forward propagation. In image description tasks, visual attention is involved to highlight which regions of the image that the model should emphasize \cite{song2018pixels}, and it is also applicable in video captioning which combines with neural networks to identify salient frames that the network should pay attention to \cite{gao2017video}. 

Adding attentions in action streams can facilitate exploring motion saliency through stripping background information \cite{sharma2015action}, exploiting pose attention from human actions \cite{du2017rpan}, or assigning more weights to engaged joints and active frames in 3D skeleton dynamics \cite{liu2018skeleton}. This comes from the fact that, for example, if one character is moving his or her arm towards another character, we need to lock the arm movement of the compelling character and react accordingly. However, if one character approaches another character with a kick, then we may focus on the active leg and dodge at an appropriate timestamp. In synthesizing interactions, \cite{baruah2020multimodal} attended to the informative joints to synthesize the reactive features which motivates our work to explore the synchronization of the two characters during the interaction.

\section{Preliminaries}
\label{sec:preliminaries}
\subsection{Generative Adversarial Networks}
\label{sec:2.1}
Generative adversarial networks (GAN) \cite{goodfellow2014generative} is introduced from game theory that a generator and a discriminator contrast with each other to achieve a Nash equilibrium \cite{salimans2016improved}. The generative model $G$ processes a random variable $z$ to $G(z)$ which will be evaluated by the discriminative model $D$, and the function of $D$ is to differentiate the real sample $x$ from the fake sample $G(z)$. The objective function of training GAN follows a minimax optimization procedure:
\begin{multline}
\min\limits_{G}\max\limits_{D}L_{GAN} (G,D)=\\
\mathbb{E}_{x\sim p_{data}(x)}[\log D(x)]+\mathbb{E}_{z\sim p_{z}(z)}[\log (1-D(G(z)))]
\end{multline}

With GAN and its vast variations, one can generate vivid samples such as images \cite{liu2016coupled} or videos \cite{xu2019prediction} following real-world data distributions judged by the discriminator. In this paper, we utilize the power of a binary and a multi-class discriminator to enhance the quality of the synthesized reactive motion with realistic and discriminative dynamics.

\subsection{Seq2seq Attention Mechanism}
\label{sec:2.2}
The seq2seq attention~\cite{bahdanau2014neural} aims to establish a bridge between encoder and decoder to emphasize the informative steps and improve output quality in decoding. Specifically, with a RNN-based backbone, seq2seq attention at each decoder step $t$ learns a context vector $r_{t}$ from the weighted summation of all the encoded states $\{h_s\}_{s=1}^S$ by:
\begin{equation}
r_{t}=\sum_{s=1}^{S} \alpha(s,t)h_{s}.\label{eq:context}
\end{equation}
Here, the attention weight $\alpha(s,t)$ is a content-based addressing function that evaluates the general score between encoder state $h_{s}$ and the previous decoder state $\hat{h}_{t-1}$ given by:
\begin{equation}
\alpha(s,t)=softmax(Vtanh(W[h_{s};\hat{h}_{t-1}])),\label{eq:addressing}
\end{equation}
where $W$ is a fully connected matrix to keep the dimension consistent. The seq2seq attention can be either global or local depending on whether all or a part of the hidden states of the encoder are included~\cite{luong2015effective}.

Since using global attention in a seq2seq architecture can effectively model the dependencies between the input dynamics and the previous decoder step, in this paper, we adapt it to strengthen the stepwise correlations between two characters in an interaction. 



\begin{figure*}[!t]
\centering\includegraphics[width=.98\linewidth]{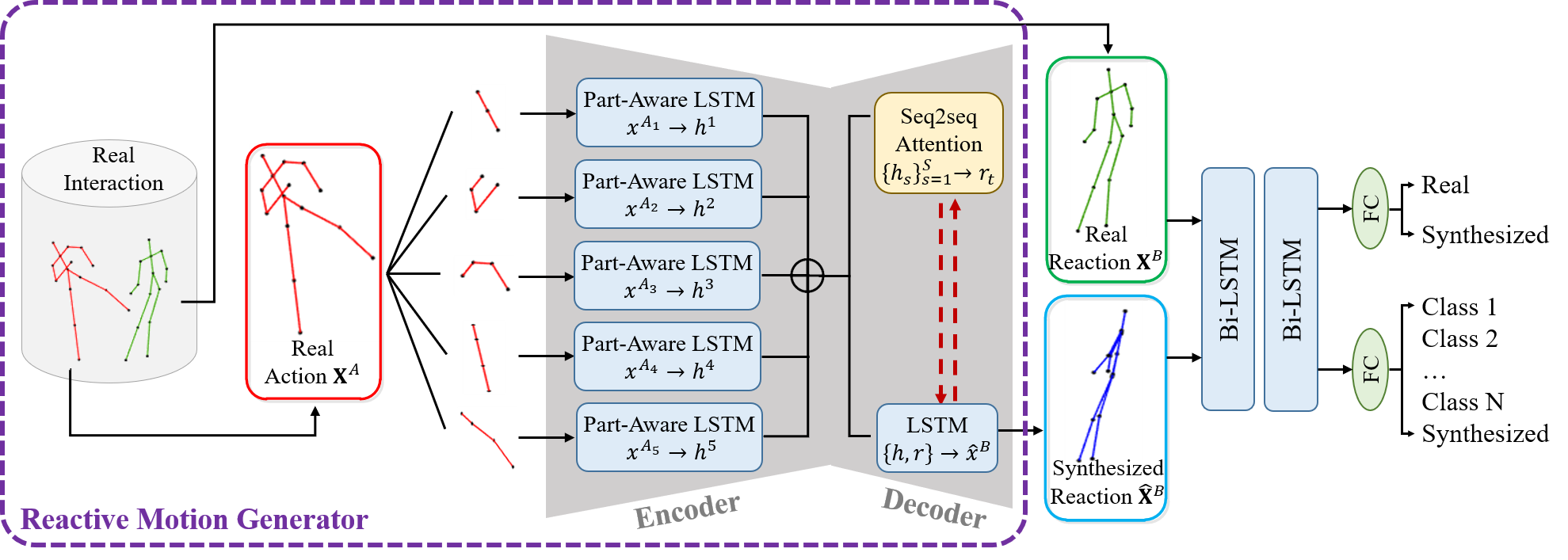}
\caption{An overview of the proposed reaction generation architecture.}
\label{fig:1} 
\end{figure*}

\section{Reactive Motion Synthesis}
\label{sec:method}

In an interaction involving two characters denoted as A and B, we consider character A to be the one performing the intended action, and character B to be the one reacting. The aim of our system is to synthesize the motion of B given that of A. As data pre-processing, we normalize the interaction by rotating them according to the facing direction of A, and translating the origin point of the new coordinate system to the pelvis joint of A. B's joint locations are then represented under such a transformation. 

The framework of our reaction generation can be found in Fig.~\ref{fig:1}. The overall network is trained by integrating three auxiliary constraints: bone, continuity and contractive losses, that target at reinforcing the adversarial objective with physical properties, stability and continuity of the synthesized motion sequence, respectively. The architecture of our reactive motion synthesis system consists of two parts: a part-based attentive recurrent generator to synthesize reaction from the input action, and a dual discriminator to increase the generator capacity with type-specific realistic reactive features. 

\subsection{The Part-based Attentive Recurrent Generator}
\label{sec:3.1}
We propose a generator that synthesizes the reactive motion in an interaction. The generator does not require the class label of the interaction to be explicitly defined, which enhances the usability of the system as an animation system, since the nature of the interaction may be unclear to the animators in some scenarios. Instead, we only take in the action from the active character as the input.



We construct a part-aware recurrent generator with seq2seq attention to learn the dynamic mapping between the input and its reactive motion. For encoding the observed motion, we break down the character and separately model the body part-level dynamics. The obtained hierarchical information helps the synthesized character to better observe local movements and react properly. For generating the reactive motion, we construct an attentive LSTM decoder to temporally align the decoded reactive motion with the input character by recognizing the informative encoder steps. The part-aware encoder and attentive decoder together form our reactive motion generator $G$.

We first adopt hierarchical part-based LSTM blocks to shape the temporal variations of each input body part. With the articulated structure, human joints can be segmented into five main parts (four limbs and the trunk) \cite{si2018skeleton}. \textcolor{black}{In particular, our input and output actions are represented with 3D joint positions in Cartesian coordinate system, and we denote an interaction after normalization with $S$ frames of poses as: $\{\mathbf{X}^{A},\mathbf{X}^{B}\}=\{(x_{s}^{A},x_{s}^{B})\}_{s=1}^{S}=\{(x_{s}^{A_{p}},x_{s}^{B_{p}})\}_{s,p=1,1}^{S,5}$
with the body part index $p$.} 
In the encoder, the LSTM neuron takes $x_{s}^{A_{p}}$ of character A at frame $s$ as the input to generate the hidden state $h_{s}^{p}$, and its previous state of the decoder $h_{s-1}^{p}$ is also participated in each LSTM cell to update the input gate $i_{t}^{p}$, the output gate $o_{t}^{p}$, the forget gate $f_{t}^{p}$, the interim gate $u_{t}^{p}$, and the cell gate $c_{t}^{p}$ for the $p$-th body part respectively by the equations:
\begin{equation}
\begin{pmatrix}i_{s}^{p}\\ f_{s}^{p}\\o_{s}^{p}\\u_{s}^{p}\end{pmatrix}=\begin{pmatrix}\sigma\\ \sigma\\\sigma\\\tanh\end{pmatrix}W_{p}\begin{pmatrix}x_{s-1}^{A_{p}}\\\\\\ h_{s-1}^{p}\end{pmatrix},
\end{equation}
\begin{equation}
c_{s}^{p}=f_{s}^{p}\odot c_{s-1}^{p}+i_{s}^{p}\odot u_{s}^{p},\label{eq:cell}
\end{equation}
\begin{equation}
h_{s}^{p}=o_{s}^{p}\odot\tanh(c_{s}^{p}),\label{eq:hidden}
\end{equation}
where $W_{p}$ represents the shared LSTM weights for all the joints in the $p$-th body part. Then, the five local hidden states go through a concatenated layer to formulate the final integrated spatial state $h_{s}=h_{s}^{1}\oplus\ldots\oplus h_{s}^{5}$ of the whole body, which can be regarded as a precise geometric refinement at the $s$ frame step.

In our decoder phase, the attention mechanism introduced in Sect.~\ref{sec:2.2} is integrated with a LSTM layer to focus on the crucial information among rich temporal data for each decoder state $\hat{h}_{t}$. The context vector $r_{t}$ obtained from the probability combination of all the hidden states in the connected hierarchical-LSTM layer is calculated by Equations (\ref{eq:context}) and (\ref{eq:addressing}), and then $r_{t}$ is used to update all the potential gates of the LSTM decoder at step $t$ as well as the motion output $\hat{x}_{t}^{B}$ with attention significance:
\begin{equation}
\begin{pmatrix}\hat{i}_{t}\\\hat{f}_{t}\\\hat{o}_{t}\\\hat{u}_{t}\\\hat{x}_{t}\end{pmatrix}=\begin{pmatrix}\sigma\\ \sigma\\\sigma\\\tanh\\\tanh\end{pmatrix}\hat{W}\begin{pmatrix}\hat{x}_{t-1}^{B}\\\\ \hat{h}_{t-1}\\\\r_{t}\end{pmatrix}.
\end{equation}
where $\hat{c}_{t}$ and $\hat{h}_{t}$ are updated using the same configuration as in (\ref{eq:cell}) and (\ref{eq:hidden}). Since the generated motion for character B should have the same number of frames as the input motion for character A to complete an effective interaction, $S$ and $T$ are set to be equal in our encoder-decoder model. Besides, we constructed a linear layer after the attentive LSTM layer to restore the reactive pose at each timestep $t$. 

\begin{figure*}[t]
\centering\includegraphics[width=.7\linewidth]{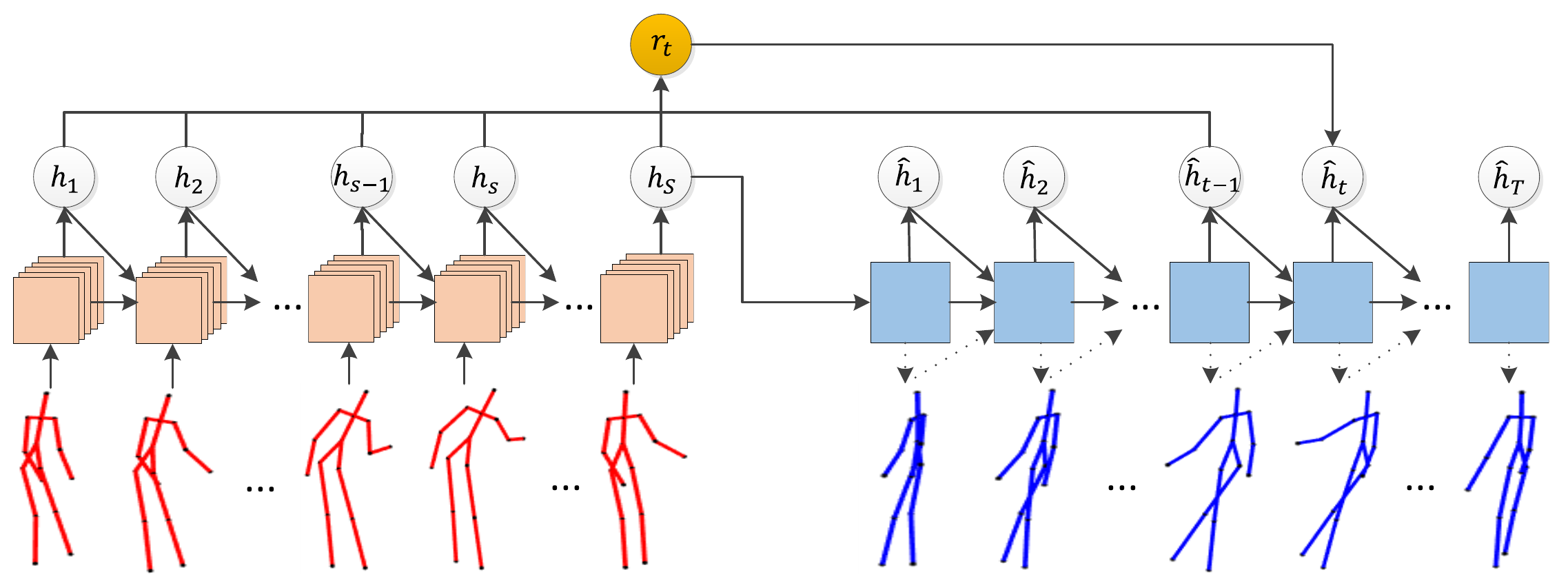}
\caption{The reactive motion generator pipeline. The characters in red show example frames of real-world shaking hand and the blue characters are example frames of the synthesized reaction.}
\label{fig:2} 
\end{figure*}

We attach the attentive layer to help strengthen the correlations between the encoder and decoder by informing the importance of all the encoder steps to the current decoder step. With an effective context vector linking the encoder and decoder per frame, the attentive mechanism brings an actual effect that temporally aligns the synthesized reactive B with the observed A. The detailed attention-based generator is illustrated in Fig.~\ref{fig:2}. 

\subsection{The Class-aware Discriminator}
\label{sec:3.2}

We propose a two-way discriminator that not only identifies natural reactions $x_{B}$ from the synthesis $\hat{x}_B$, but also classifies which interaction type it belongs to. This is driven by the observation that the reactive motion of different types of interactions can be significantly different. Being able to tell the class of the interaction helps increase the capacity of the generator by synthesizing high-quality reactions with diverse reactive patterns.

We present a dual discriminator structure, in which we construct a standard binary classifier $D_{b}$ to maintain the authenticity, and a multi-class classifier $D_{m}$ to promote the discriminability of the synthesis. With the assistance of $D_{m}$, we can prevent $G$ from creating monotonous reactions for all kinds of input actions, while preserving the natures learned from the class-specific information to build a desired yet precise representation to react. As shown in the right part of Fig.~\ref{fig:1}, since most of the structures are shared between $D_{b}$ and $D_{m}$, the dual discriminator is efficient without introducing massive extra parameters to learn.

To avoid abuse of the input motion, we only feed in the synthesized reactive motion to the dual discriminator. This is because if both the real A and synthesized B are visible, the discriminator will mainly rely on extracting features from the input A for classification. As a result, less effective features are learned to justify the reactive motion that will ultimately downgrade the ability of the discriminator. On the contrary, only observing the movement of character B will enforce the discriminator focusing on the reactive pattern to increase its discriminability. 

Specifically, we consider bidirectional LSTM layers shared between the two classifiers in the dual discriminator to globally execute the reactive dynamics, each of which will further go through a fully connected layer to achieve the two classification tasks, respectively. Since for the discriminator architecture, empirically under a bidirectional procedure, exploiting contextual information from both the forward and backward movements can summarize high-level features that significantly boosts the classification performance compared with its undirected counterpart \cite{zhang2015bidirectional}.

\begin{figure*}
\centering
    \begin{minipage}{0.08\textwidth}
    (a)\quad\rotatebox[origin=c]{90}{Kick}
    \end{minipage}
    \begin{minipage}{0.9\textwidth}
\includegraphics[width=.09\linewidth]{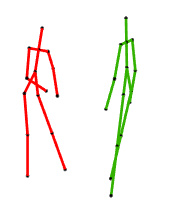}
  \hfill
  \includegraphics[width=.09\linewidth]{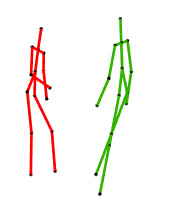}
  \hfill
  \includegraphics[width=.09\linewidth]{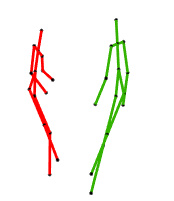}
  \hfill
  \includegraphics[width=.09\linewidth]{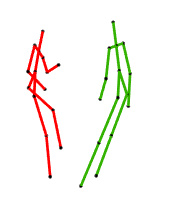}
  \hfill
  \includegraphics[width=.09\linewidth]{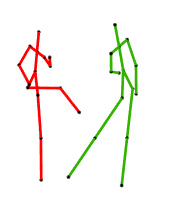}
  \hfill
  \includegraphics[width=.09\linewidth]{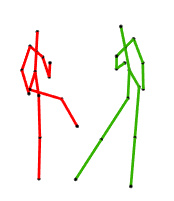}
  \hfill
  \includegraphics[width=.09\linewidth]{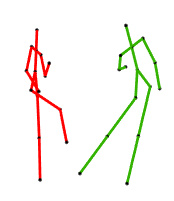}
  \hfill
  \includegraphics[width=.09\linewidth]{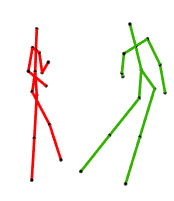}
  \hfill
  \includegraphics[width=.09\linewidth]{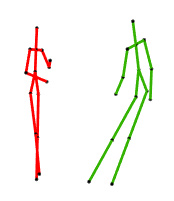}
  \hfill
  \includegraphics[width=.09\linewidth]{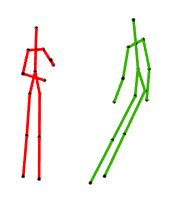}
  \hfill \\
  \includegraphics[width=.09\linewidth]{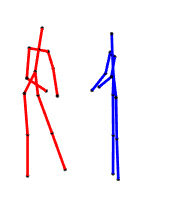}
  \hfill
  \includegraphics[width=.09\linewidth]{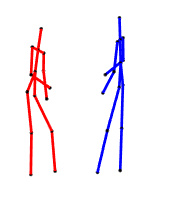}
  \hfill 
  \includegraphics[width=.09\linewidth]{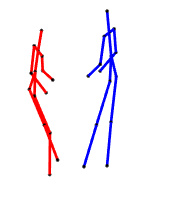}
  \hfill 
  \includegraphics[width=.09\linewidth]{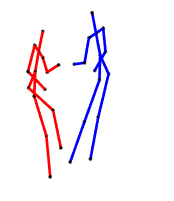}
  \hfill 
  \includegraphics[width=.09\linewidth]{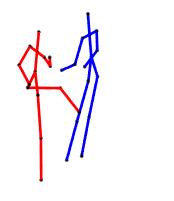}
  \hfill 
  \includegraphics[width=.09\linewidth]{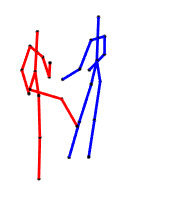}
  \hfill 
  \includegraphics[width=.09\linewidth]{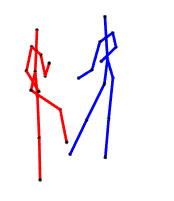}
  \hfill 
  \includegraphics[width=.09\linewidth]{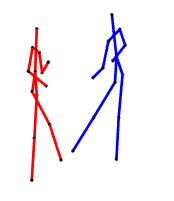}
  \hfill 
  \includegraphics[width=.09\linewidth]{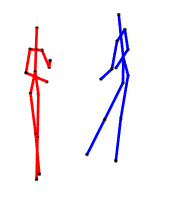}
  \hfill 
  \includegraphics[width=.09\linewidth]{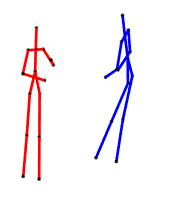}
  \hfill 
    \end{minipage}\\
    \begin{minipage}{0.08\textwidth}
       (b)\quad\rotatebox[origin=c]{90}{Push}
    \end{minipage}
    \begin{minipage}{0.9\textwidth}
  \includegraphics[width=.09\linewidth]{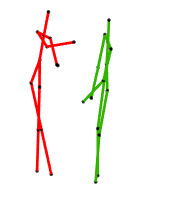}
  \hfill
  \includegraphics[width=.09\linewidth]{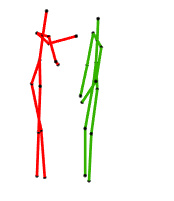}
  \hfill
  \includegraphics[width=.09\linewidth]{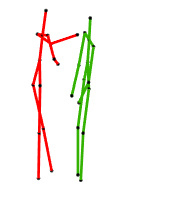}
  \hfill
  \includegraphics[width=.09\linewidth]{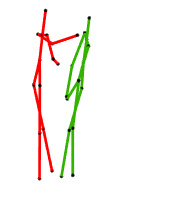}
  \hfill
  \includegraphics[width=.09\linewidth]{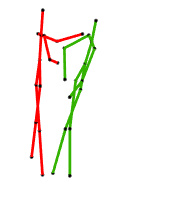}
  \hfill
  \includegraphics[width=.09\linewidth]{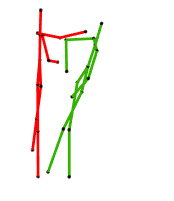}
  \hfill
  \includegraphics[width=.09\linewidth]{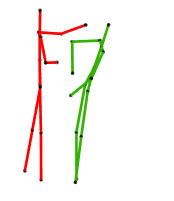}
  \hfill
  \includegraphics[width=.09\linewidth]{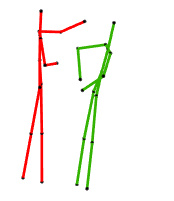}
  \hfill
  \includegraphics[width=.09\linewidth]{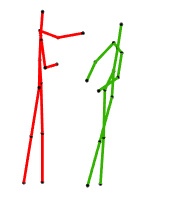}
  \hfill
  \includegraphics[width=.09\linewidth]{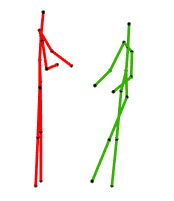}
  \hfill \\
  \includegraphics[width=.09\linewidth]{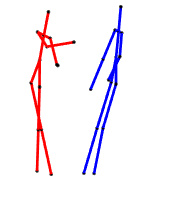}
  \hfill
  \includegraphics[width=.09\linewidth]{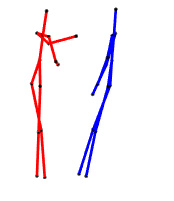}
  \hfill
  \includegraphics[width=.09\linewidth]{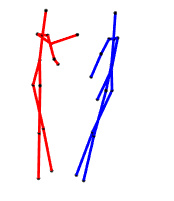}
  \hfill
  \includegraphics[width=.09\linewidth]{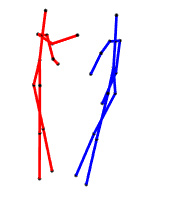}
  \hfill
  \includegraphics[width=.09\linewidth]{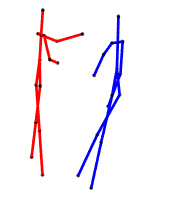}
  \hfill
  \includegraphics[width=.09\linewidth]{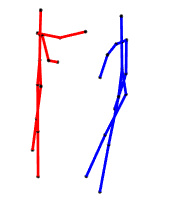}
  \hfill
  \includegraphics[width=.09\linewidth]{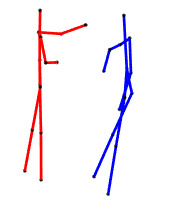}
  \hfill
  \includegraphics[width=.09\linewidth]{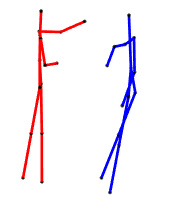}
  \hfill
  \includegraphics[width=.09\linewidth]{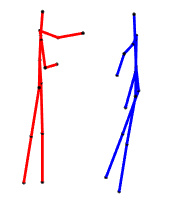}
  \hfill
  \includegraphics[width=.09\linewidth]{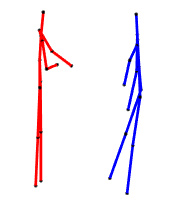}
  \hfill
    \end{minipage}\\
    \begin{minipage}{0.08\textwidth}
       (c)\quad\rotatebox[origin=c]{90}{Punch}
    \end{minipage}
    \begin{minipage}{0.9\textwidth}
  \includegraphics[width=.09\linewidth]{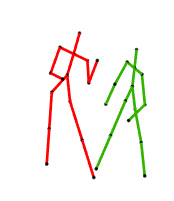}
  \hfill
  \includegraphics[width=.09\linewidth]{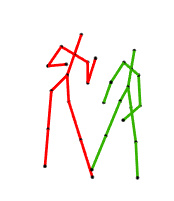}
  \hfill
  \includegraphics[width=.09\linewidth]{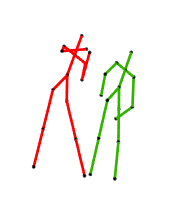}
  \hfill
  \includegraphics[width=.09\linewidth]{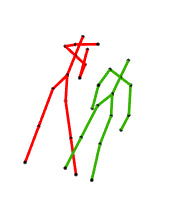}
  \hfill
  \includegraphics[width=.09\linewidth]{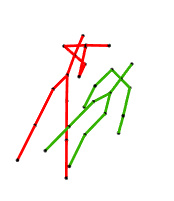}
  \hfill
  \includegraphics[width=.09\linewidth]{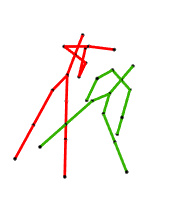}
  \hfill
  \includegraphics[width=.09\linewidth]{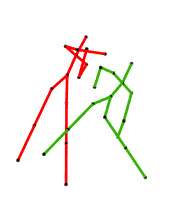}
  \hfill
  \includegraphics[width=.09\linewidth]{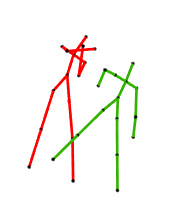}
  \hfill
  \includegraphics[width=.09\linewidth]{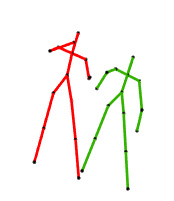}
  \hfill
  \includegraphics[width=.09\linewidth]{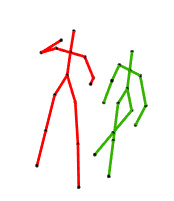}
  \hfill \\
  \includegraphics[width=.09\linewidth]{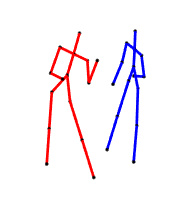}
  \hfill
  \includegraphics[width=.09\linewidth]{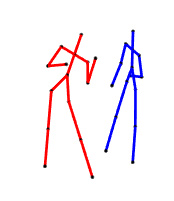}
  \hfill
  \includegraphics[width=.09\linewidth]{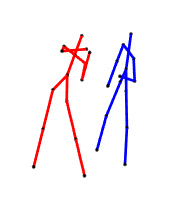}
  \hfill
  \includegraphics[width=.09\linewidth]{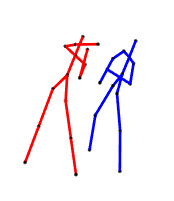}
  \hfill
  \includegraphics[width=.09\linewidth]{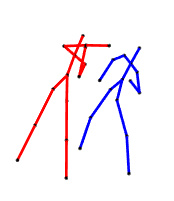}
  \hfill
  \includegraphics[width=.09\linewidth]{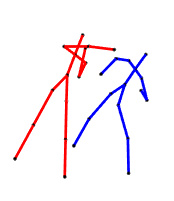}
  \hfill
  \includegraphics[width=.09\linewidth]{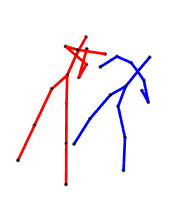}
  \hfill
  \includegraphics[width=.09\linewidth]{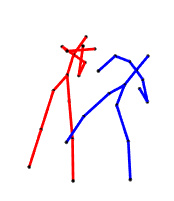}
  \hfill
  \includegraphics[width=.09\linewidth]{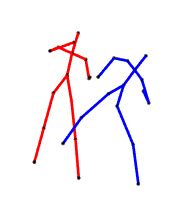}
  \hfill
  \includegraphics[width=.09\linewidth]{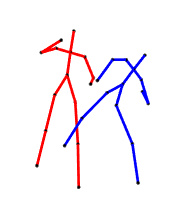}
  \hfill
    \end{minipage}\\
    \begin{minipage}{0.08\textwidth}
       (d)\quad\rotatebox[origin=c]{90}{Hug}
    \end{minipage}
    \begin{minipage}{0.9\textwidth}
  \includegraphics[width=.09\linewidth]{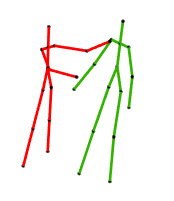}
  \hfill
  \includegraphics[width=.09\linewidth]{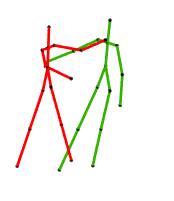}
  \hfill
  \includegraphics[width=.09\linewidth]{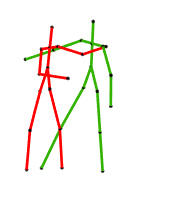}
  \hfill
  \includegraphics[width=.09\linewidth]{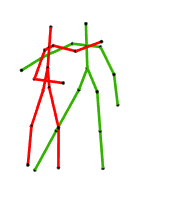}
  \hfill
  \includegraphics[width=.09\linewidth]{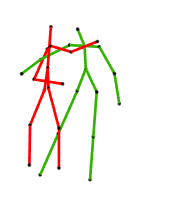}
  \hfill
  \includegraphics[width=.09\linewidth]{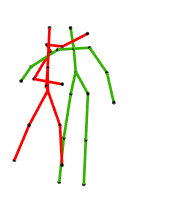}
  \hfill
  \includegraphics[width=.09\linewidth]{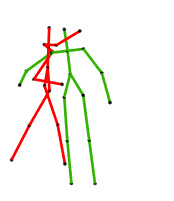}
  \hfill
  \includegraphics[width=.09\linewidth]{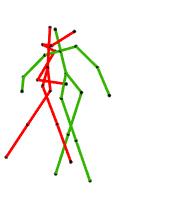}
  \hfill
  \includegraphics[width=.09\linewidth]{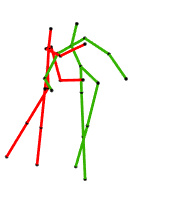}
  \hfill
  \includegraphics[width=.09\linewidth]{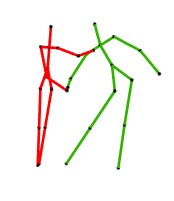}
  \hfill \\
  \includegraphics[width=.09\linewidth]{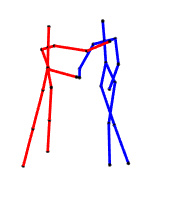}
  \hfill
  \includegraphics[width=.09\linewidth]{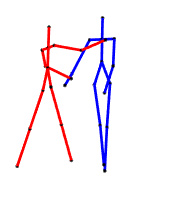}
  \hfill
  \includegraphics[width=.09\linewidth]{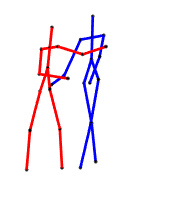}
  \hfill
  \includegraphics[width=.09\linewidth]{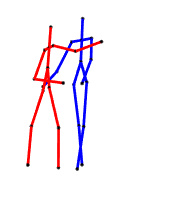}
  \hfill
  \includegraphics[width=.09\linewidth]{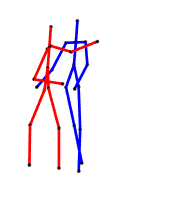}
  \hfill
  \includegraphics[width=.09\linewidth]{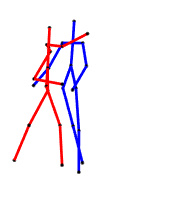}
  \hfill
  \includegraphics[width=.09\linewidth]{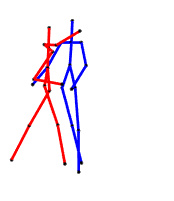}
  \hfill
  \includegraphics[width=.09\linewidth]{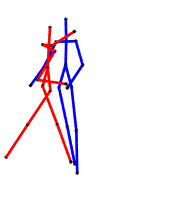}
  \hfill
  \includegraphics[width=.09\linewidth]{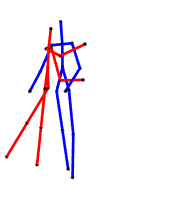}
  \hfill
  \includegraphics[width=.09\linewidth]{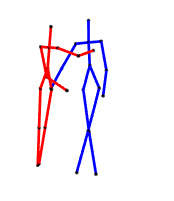}
  \hfill
    \end{minipage}\\
    \begin{minipage}{0.08\textwidth}
       (e)\quad\rotatebox[origin=c]{90}{Shake hands}
    \end{minipage}
    \begin{minipage}{0.9\textwidth}
  \includegraphics[width=.09\linewidth]{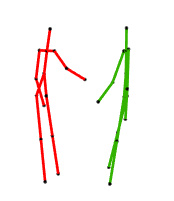}
  \hfill
  \includegraphics[width=.09\linewidth]{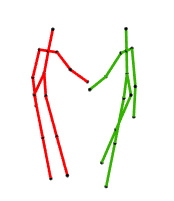}
  \hfill
  \includegraphics[width=.09\linewidth]{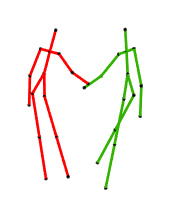}
  \hfill
  \includegraphics[width=.09\linewidth]{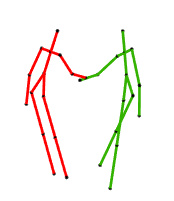}
  \hfill
  \includegraphics[width=.09\linewidth]{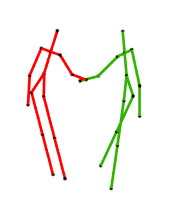}
  \hfill
  \includegraphics[width=.09\linewidth]{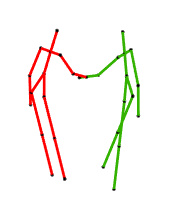}
  \hfill
  \includegraphics[width=.09\linewidth]{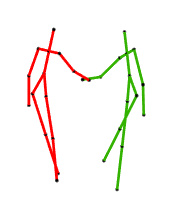}
  \hfill
  \includegraphics[width=.09\linewidth]{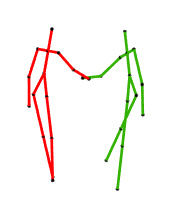}
  \hfill
  \includegraphics[width=.09\linewidth]{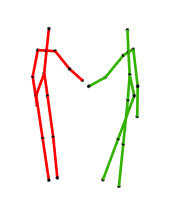}
  \hfill
  \includegraphics[width=.09\linewidth]{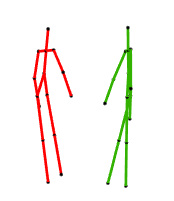}
  \hfill \\
  \includegraphics[width=.09\linewidth]{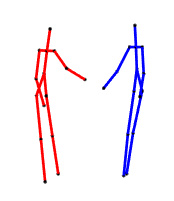}
  \hfill 
  \includegraphics[width=.09\linewidth]{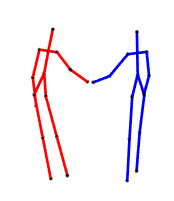}
  \hfill 
  \includegraphics[width=.09\linewidth]{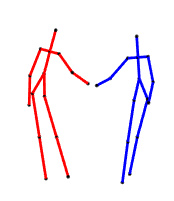}
  \hfill 
  \includegraphics[width=.09\linewidth]{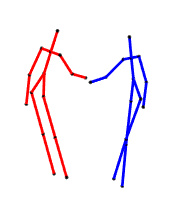}
  \hfill 
  \includegraphics[width=.09\linewidth]{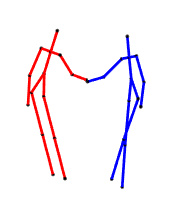}
  \hfill 
  \includegraphics[width=.09\linewidth]{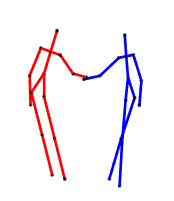}
  \hfill 
  \includegraphics[width=.09\linewidth]{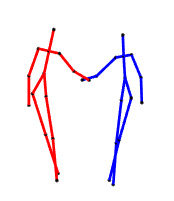}
  \hfill 
  \includegraphics[width=.09\linewidth]{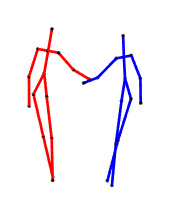}
  \hfill 
  \includegraphics[width=.09\linewidth]{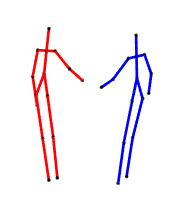}
  \hfill 
  \includegraphics[width=.09\linewidth]{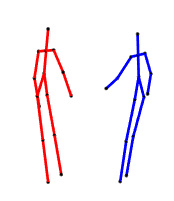}
  \hfill
    \end{minipage}\\
    \begin{minipage}{0.08\textwidth}
       (f)\quad\rotatebox[origin=c]{90}{Exchange objects}
    \end{minipage}
    \begin{minipage}{0.9\textwidth}
  \includegraphics[width=.09\linewidth]{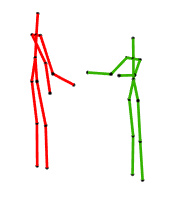}
  \hfill
  \includegraphics[width=.09\linewidth]{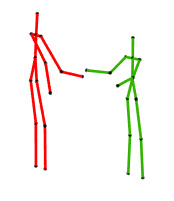}
  \hfill
  \includegraphics[width=.09\linewidth]{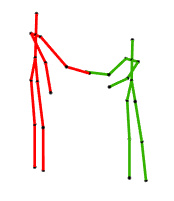}
  \hfill
  \includegraphics[width=.09\linewidth]{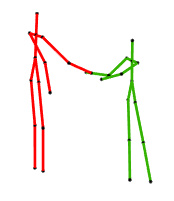}
  \hfill
  \includegraphics[width=.09\linewidth]{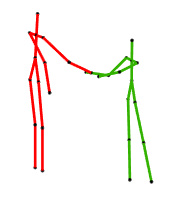}
  \hfill
  \includegraphics[width=.09\linewidth]{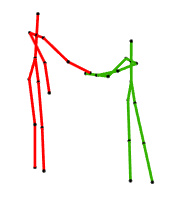}
  \hfill
  \includegraphics[width=.09\linewidth]{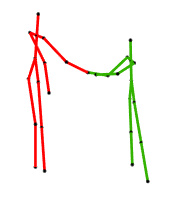}
  \hfill
  \includegraphics[width=.09\linewidth]{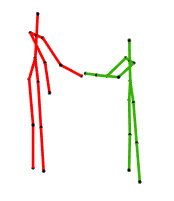}
  \hfill
  \includegraphics[width=.09\linewidth]{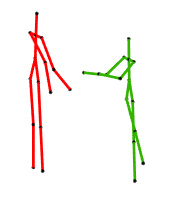}
  \hfill
  \includegraphics[width=.09\linewidth]{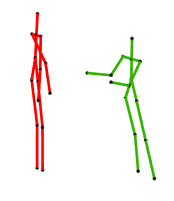}
  \hfill \\
  \includegraphics[width=.09\linewidth]{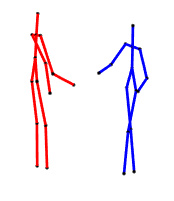}
  \hfill
  \includegraphics[width=.09\linewidth]{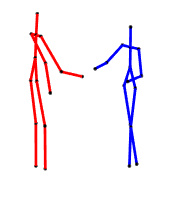}
  \hfill
  \includegraphics[width=.09\linewidth]{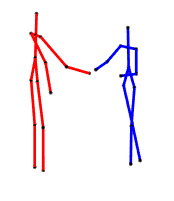}
  \hfill
  \includegraphics[width=.09\linewidth]{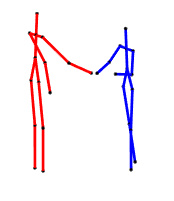}
  \hfill
  \includegraphics[width=.09\linewidth]{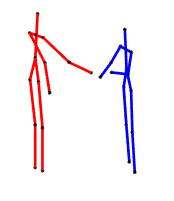}
  \hfill
  \includegraphics[width=.09\linewidth]{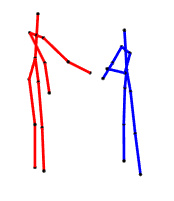}
  \hfill
  \includegraphics[width=.09\linewidth]{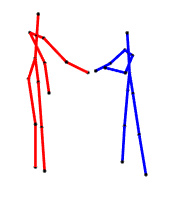}
  \hfill
  \includegraphics[width=.09\linewidth]{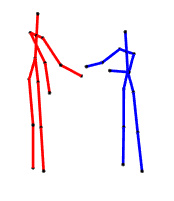}
  \hfill
  \includegraphics[width=.09\linewidth]{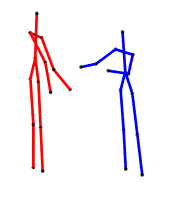}
  \hfill
  \includegraphics[width=.09\linewidth]{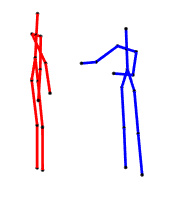}
  \hfill
    \end{minipage}\\
    \caption{\label{fig:3}The ground truth and the synthesis for SBU dataset for different classes of interactions. The red character is the observation. The green and blue characters are the ground truth and the synthesis, respectively. 
    }
\end{figure*}

\begin{figure*}[t]
\centering
    \begin{minipage}{0.08\textwidth}
        (a)\quad\rotatebox[origin=c]{90}{Kick Sample 1}
    \end{minipage}
    \begin{minipage}{0.9\textwidth}
        \includegraphics[width=1\linewidth]{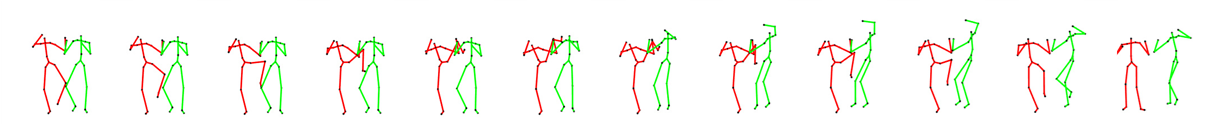}
        \includegraphics[width=1\linewidth]{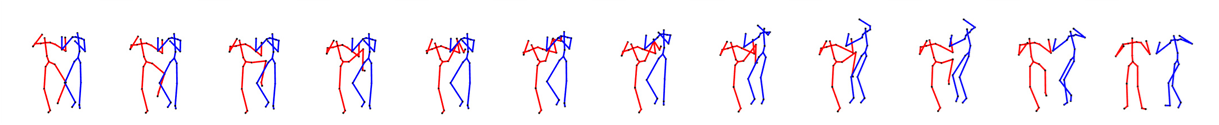}
    \end{minipage}
    \begin{minipage}{0.08\textwidth}
        (b)\quad\rotatebox[origin=c]{90}{Kick Sample 2}
    \end{minipage}
    \begin{minipage}{0.9\textwidth}
        \includegraphics[width=1\linewidth]{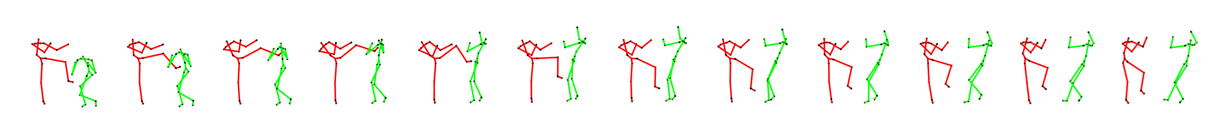}
        \includegraphics[width=1\linewidth]{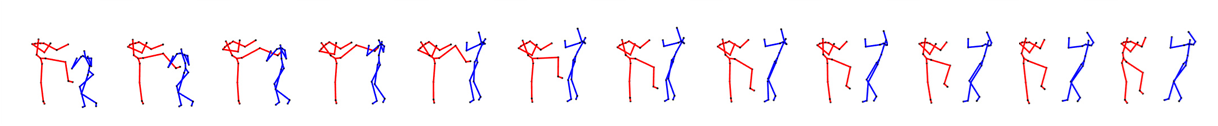}
    \end{minipage}
    \begin{minipage}{0.08\textwidth}
        (c)\quad\rotatebox[origin=c]{90}{Punch Sample 1}
    \end{minipage}
    \begin{minipage}{0.9\textwidth}
        \includegraphics[width=1\linewidth]{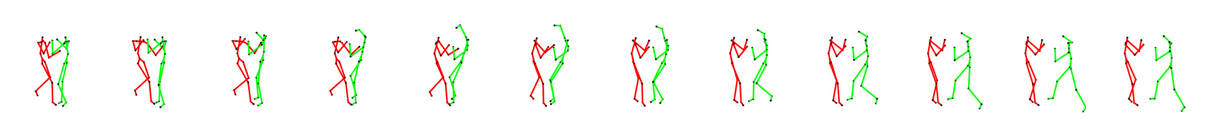}
        \includegraphics[width=1\linewidth]{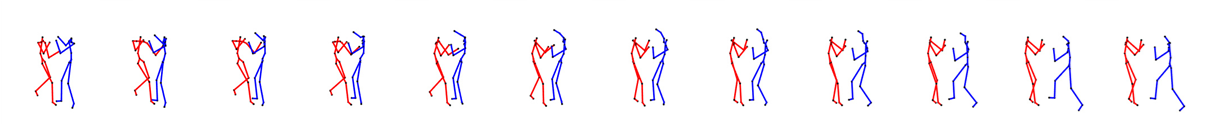}
    \end{minipage}
    \begin{minipage}{0.08\textwidth}
        (d)\quad\rotatebox[origin=c]{90}{Punch Sample 2}
    \end{minipage}
    \begin{minipage}{0.9\textwidth}
        \includegraphics[width=1\linewidth]{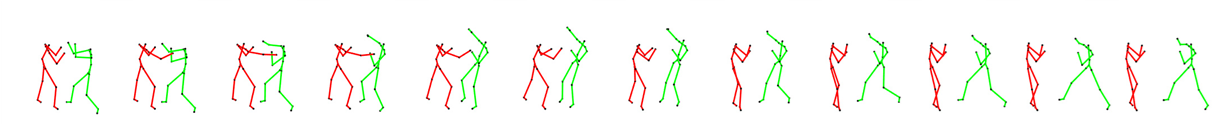}
        \includegraphics[width=1\linewidth]{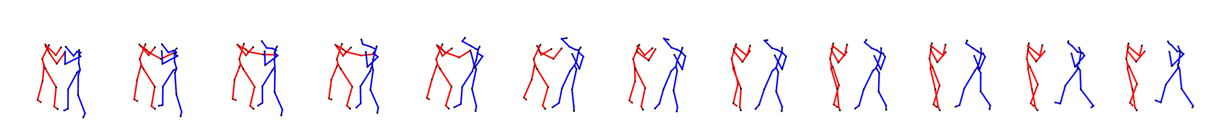}
    \end{minipage}\\
    \caption{\label{fig:2C}\textcolor{black}{The ground truth and the synthesis for the high-quality 2C dataset for \emph{kicking} and \emph{punching}. The red character is the observation. The green and blue characters are the ground truth and the synthesis, respectively.} 
    }
\end{figure*}

\subsection{The Loss Functions}
\label{sec:3.3}

The adversarial system of our reactive motion generator and the class-aware discriminator is trained based on a semi-supervised loss inspired by \cite{salimans2016improved}. Traditionally, the aim of semi-supervised GAN \cite{salimans2016improved,odena2016semi, kumar2017semi} is to learn a capable classifier that can recognize real samples. In contrast, we utilize the classification ability of the multi-class classifier to generate samples of different classes, such that the generator can learn from the class-specific information to synthesize a better reaction.

Our multi-class classifier $D_{m}$ is supervised for discriminating whether a reactive motion belongs to any of the real $N$ classes or the fake class $N+1$, and our binary classifier $D_{b}$ is unsupervised that tells the real reaction from fake. The overall semi-supervised adversarial loss can thus be expressed by the supervised $\mathcal{L}_{sup}$ and unsupervised $\mathcal{L}_{unsup}$ components as:
\begin{multline}
\mathcal{L}_{sup}=-\mathbb{E}_{x,y\sim G}\log\frac{p_{D_{m}}(y|x,y<N+1)}{p_{D_{m}}(y|x,y=N+1)}\\
+\mathbb{E}_{x,y\sim p_{B}}\log p_{D_{m}}(y|x,y<N+1),\label{eq:sup}
\end{multline}
\begin{equation}
\begin{split}
\mathcal{L}_{unsup}&=\mathbb{E}_{x\sim p_{B}}\log [1-p_{D_{b}}(y_{syn}|x)]+\mathbb{E}_{x\sim G}\log p_{D_{b}}(y_{syn}|x)\\
&=\mathbb{E}_{x\sim p_{B}}\log D_{b}(x)+\mathbb{E}_{x\sim p_{A}}\log [1-D_{b}(G(x))].\label{eq:unsup}
\end{split}
\end{equation}
where $p_{A}$ and $p_{B}$ stand for the real data distributions of the motions from character A and B, respectively, $y$ is the class label for the input action $x$ and $p(y_{syn}|x)$ represents the probability of $x$ being classified as the synthesized class. In $\mathcal{L}_{unsup}$, we denote $D_{b}(x)=1-p_{D_{b}}(y_{syn}|x)$ so that it can be rewritten into the form of standard objective function of GAN. 

Different from the normal semi-supervised GAN, our multi-class classifier $D_{m}$ also classifies the synthesized reaction. This is done by employing a new term $\mathbb{E}_{x,y\sim G}\log\frac{p_{D_{m}}(y|x,y<N+1)}{p_{D_{m}}(y|x,y=N+1)}$ to the supervised $D_{m}$.
Compared to conditional GAN \cite{mirza2014conditional}, we do not adopt label information into the generator but only for the discriminator, since our generator will create plausible responses that can be recognized as the underlying interaction type without early annotation.

We further design three loss functions for synthesizing high-quality movement as follows:

\textbf{Bone loss:} To synthesize a valid motion, it is essential to preserve bone lengths among all the generated frames, and we use an additional loss function $\mathcal{L}_{skl}$ to restrict this physical constraint:
\begin{equation}
\mathcal{L}_{skl}=\sum_{t}\sum_{j}\big\vert skl(\hat{x}_{t}^{B},j)-skl_{ref}(j)\big\vert\label{eq:bone},
\end{equation}
where $skl(\hat{x}_{t}^{B},j)$ is the predicted skeleton length at time $t$ and $skl_{ref}(j)$ is the reference skeleton length with $j$ denoting the bone index. The ground truth skeleton length $skl(x_{t}^{B},j)$ is character specific so a uniform constant $skl_{ref}(j)$ is used instead, as the intention of our network is not to shape the physiological properties (e.g. bone length, height) of the people in front, 
but to predict the tendency of motion kinetics.

\textbf{Continuity loss:} Similar to \cite{yan2017skeleton} that designs a triple loss to maintain video appearance consistency based on pixel difference, we demonstrate the continuity loss based on joint locations, which is beneficial to synthesize smooth and stable motion. The modified continuity loss for skeleton-based motion sequence is defined as:
\begin{equation}
\mathcal{L}_{con}=\sum_{t}\max(\vert \Vert \hat{x}_{t+\Delta t}^{B}-\hat{x}_{t}^{B}\Vert^{2}-\Vert \hat{x}_{t+k\Delta t}^{B}-\hat{x}_{t}^{B}\Vert^{2}+\lambda\vert,0),
\end{equation}
where $\Delta t$ is temporal gradient and $\lambda$ measures the sensitiveness of the constructed activity. A small $\lambda$ demands to narrow the gap between close frames (differ by $\Delta t$) and remote ones (differ by $k\Delta t$) to obtain a smooth motion. By tuning the intrinsic parameters $\lambda$, $\Delta t$ and $k$, we can control the quantity of random movements emerged in $\mathbf{\hat{X}}^{B}$.

\textbf{Contractive loss:} We also adopt the $L_{1}$ norm for training the generator to make sure it follows the real reactive patterns, which will also strongly guide the reactive movements and reduce ambiguous predictions. 
Therefore, a contractive loss under $L_{1}$ norm is formulated to approximate the ground truth reaction:
\begin{equation}
\mathcal{L}_{1}=\sum_{t}\vert \hat{x}_{t}^{B}-x_{t}^{B}\vert.
\end{equation}
This loss aims to mimic specific motion style to avoid neutrality and monotonous generation. 

The overall min-max objective function of the reaction generation architecture is the combination of all the network losses:
\begin{equation}
\min\limits_{G}\max\limits_{D}\mathcal{L}_{sup}+\mathcal{L}_{unsup}+\alpha\mathcal{L}_{skl}+\beta\mathcal{L}_{con}+\gamma\mathcal{L}_{1},\label{eq:overall}
\end{equation}
where $\alpha$, $\beta$ and $\gamma$ control the weights of the respective losses.

\section{Experimental Results}
\label{sec:experiment}

\textbf{Dataset settings:} To demonstrate the effectiveness of our approach on 3D joint space, we evaluate on both Kinect-based datasets, i.e. SBU Kinect Interaction dataset (SBU) \cite{yun2012two} and Human-Human-Object Interaction dataset (HHOI) \cite{shu2016learning}, and high-quality Motion Capture-based Character-Character dataset (2C)~\cite{shen2019interaction}. The SBU dataset includes 8 interaction categories (i.e., \emph{approach}, \emph{depart}, \emph{kick}, \emph{push}, \emph{punch}, \emph{hug}, \emph{shake hands} and \emph{exchange objects}) performed by 7 participants. It also provides the annotations of ``active" agent (character A) and ``inactive" agent (character B). 
We exclude \emph{approach} and \emph{depart} since in these interactions the indicated character stands still and no movement is presented for forecasting. For HHOI dataset, we experiment on 2 types of human-human interactions: \emph{shake hands} and \emph{high-five}. Compared with SBU dataset, HHOI contains fewer instances in each category but a longer duration with more frames in each captured sequence. To better fit the network, we expand the dataset by clipping a sliding window with the size of 40 frames and shifting every 5 frames along the sequence. On both datasets, we conduct leave-one-subject-out cross-validation. 
\textcolor{black}{The 2C dataset contains \emph{kicking} and \emph{punching} interactions with about 50 clips in total. In this high-quality dataset, each character contains 20 joints and we convert the 3D joint angle representations into joint positions using forward kinematics.}

\textbf{Implementation details:} Our reaction generator is built upon the Keras platform with the TensorFlow backend. RMSprop is adapted as the optimizer with the learning rate of 0.01. There are 40 and 60 LSTM neurons for each spatial slice, and 200 and 300 for the temporal attentive layer for SBU and HHOI, respectively. \textcolor{black}{For 2C, the LSTM neurons are set to 200 and 1000 for the body slice and the attentive layer, respectively. The parameters $k$, $\Delta t$, and $\lambda$ are set to 1, 5, and 0.1, respectively. The training time is about 9.3s for each epoch and our model normally converges around 1000 epochs. The inference time for each interaction is around 5.2ms.} For the weights of network losses, we set $\alpha=\beta=0.01$, and $\gamma=1$ in Equation (\ref{eq:overall}). Since the function of $L_{skl}$ and $L_{con}$ is to prevent the abuse of physical properties, i.e., skeleton length and action smoothness, lower weights are assigned to these losses. Otherwise, the model will vacillate among various body shapes and not converge. For the adversarial loss, we also adopt one-side label smoothing \cite{salimans2016improved} to help train the discriminator. 

\subsection{Qualitative Evaluations}
\label{sec:4.1}
We demonstrate that our system can generate realistic reactive motion. Given the observation of an intentional action, the proposed mechanism can forecast the natural response which is successive in both space and time. Some example comparisons between ground truth and synthesis are visualized in Fig.~\ref{fig:3}. The synthesized character will learn from the skeletal positions and temporal synchronization for a reaction, which imitates how a human perceives an action and behaves accordingly. For example, the synthesized characters can move backward to dodge in \emph{punching}, \emph{pushing} and \emph{kicking}. Our model can also recognize the attack from different directions performed by different body parts. As in \emph{punching} and \emph{pushing}, the synthesized character leans back its upper body to avoid the arm from the observed actor. In \emph{kicking}, the synthesized character escapes the offensive leg by pulling back his lower body. In the neutral interactions (i.e., \emph{hugging}, \emph{shaking hands}, and \emph{exchanging objects}), the relative distance between two characters is first shortened then enlarged compared with the other three aggressive interactions showing a consistent increasing distance. This is because the $D_{m}$ classifier promotes the quality of the synthesized reactive motion by adding more discriminative details in each of the ground truth classes.

We also observe that in some unusual situations, the ground truth reactive motion is noisy with flickering joints due to occlusion. Our system synthesized a more natural reactive motion than the ground truth but with similar key features. This indicates that the generator we developed generalize well to model human movement. 

\textcolor{black}{Since the ground truth movements in the Kinect-based dataset (SBU) are very likely to present noisy joints and unnatural configurations, we also test the feasibility of our method on high-quality precise interactions (i.e. 2C) to remove the inherited noise from the low-quality motion data. We give example interactions with key frames showing the real and the generated reactive motions in Fig.~\ref{fig:2C}. We first observe that the synthesized reaction is highly consistent with the ground truth with natural arm and leg movements. The motion details are also sufficiently preserved in the synthesized reaction. For example, we can simulate the state from squat to stand at beginning of the reaction as shown in Fig.~\ref{fig:2C}(b). Furthermore, in the \emph{punching} of Fig.~\ref{fig:2C}(c), the necessary body contact is preserved with the punching hand of A hitting the upper body before the step back of B in the initial poses. The readers are referred to the supplementary video for more results.}

\begin{table}
\small
\centering
\caption{The effectiveness of $D_m$ evaluated with AFD on each interaction category of SBU.}
\label{tab:afd_dm}
\begin{tabular}{|l|c|c|}
\hline
AFD ($\downarrow$)            & w/o $D_m$ & w/ $D_m$ (Ours) \\ \hline
Kick            & 0.58   & \textbf{0.53}         \\ 
Push            & 0.52   & \textbf{0.52}         \\ 
Punch           & \textbf{0.44}   & 0.45         \\ 
Hug             & 0.81   & \textbf{0.72}         \\ 
Shake hands     & 0.50   & \textbf{0.44}         \\ 
Exchange object & 0.49   & \textbf{0.45}         \\ \hline
\end{tabular}
\end{table}

\begin{table}
\centering
\caption{\textcolor{black}{The effectiveness of the proposed reactive synthesis method over existing models evaluated with AFD on each interaction category of SBU.}}
\label{tab:afd}
\resizebox{\columnwidth}{!}{
\begin{tabular}{|l|c|c|c|c|c|c|}
\hline
AFD ($\downarrow$)            & NN & HMM & DMDP & KRL & ME-IOC &  Ours \\ \hline
Kick            & 0.81 & 0.92 & 0.65 & 0.92 & 0.67 & \textbf{0.53}      \\ 
Push            & 0.51 & 0.60 & \textbf{0.45} & 0.61 & 0.48 & 0.52        \\ 
Punch           & 0.56 & 0.66 & 0.48 & 0.66 & 0.52 & \textbf{0.45}        \\ 
Hug             & 0.61 & 0.67 & \textbf{0.48} & 0.81 & 0.47 & 0.72        \\ 
Shake hands     & 0.48 & 1.41 & 0.42 & 0.54 & \textbf{0.42} & 0.44       \\ 
Exchange object & 0.63 & 3.84 & 0.53 & 0.74 & 0.54 & \textbf{0.45}        \\ \hline
\end{tabular}}
\end{table}

\begin{table*}[t]
\small
\centering
\caption{Recognition performance (SBU) on the prototype and synthesized interactions on ablation study of losses .}
\label{tab:1}
\begin{tabular}{|c|l|l|c|c|c|c|c|c|c|c|}
\hline
\multicolumn{2}{|c|}{Accuracy}               & \emph{prototype}       & \multicolumn{2}{c|}{$Adv.$}                        & \multicolumn{2}{c|}{$Adv.+\mathcal{L}_{skl}$} & \multicolumn{2}{c|}{$Adv.+\mathcal{L}_{skl}+\mathcal{L}_{con}$} & \multicolumn{2}{c|}{$Adv.+\mathcal{L}_{skl}+\mathcal{L}_{con}+\mathcal{L}_{1}$} \\ \hline
\multirow{3}{*}{Aggresive} & Kick            & \textit{0.9698} & 0.8413          & \multirow{3}{*}{0.6413}          & 0.8841        & \multirow{3}{*}{0.7273}       & 0.9016                     & \multirow{3}{*}{0.7196}            & \textbf{0.9365}                & \multirow{3}{*}{\textbf{0.7921}}               \\
                           & Push            & \textit{0.8806} & 0.5755          &                                  & 0.6478        &                               & 0.7135                     &                                    & \textbf{0.7573}                &                                                \\
                           & Punch           & \textit{0.8583} & 0.5071          &                                  & 0.65          &                               & 0.5437                     &                                    & \textbf{0.6825}                &                                                \\ \hline
\multirow{3}{*}{Neutral}   & Hug             & \textit{0.8857} & \textbf{0.2381} & \multirow{3}{*}{\textbf{0.4379}} & 0.0778        & \multirow{3}{*}{0.3848}       & 0.1683                     & \multirow{3}{*}{0.4352}            & 0.2087                         & \multirow{3}{*}{0.3997}                        \\
                           & Shake hands     & \textit{0.7092} & 0.6495          &                                  & 0.6546        &                               & \textbf{0.7138}            &                                    & 0.4735                         &                                                \\
                           & Exchange object & \textit{0.81}   & 0.4261          &                                  & 0.4219        &                               & 0.4236                     &                                    & \textbf{0.5168}                &                                                \\ \hline
\end{tabular}
\end{table*}

\begin{table*}[]
\small
\centering
\caption{Recognition performance (HHOI) on the prototype and synthesized interactions on ablation study of losses.}
\label{tab:2}
\begin{tabular}{|l|c|c|c|c|c|}
\hline
Accuracy                & \emph{prototype}              & $Adv.$ & $Adv.+\mathcal{L}_{skl}$ & $Adv.+\mathcal{L}_{skl}+\mathcal{L}_{con}$ & $Adv.+\mathcal{L}_{skl}+\mathcal{L}_{con}+\mathcal{L}_{1}$ \\ \hline
High-five   & \textit{0.9785} & 0.5171      & \textbf{0.9901}  & 0.9067                & 0.9473                   \\ 
Shake hands & \textit{0.9778} & 0.9533      & 0.7132           & 0.8966                & \textbf{0.9673}          \\ 
Average     & \textit{0.9782} & 0.7352      & 0.8517           & 0.9017                & \textbf{0.9573}          \\ \hline
\end{tabular}
\end{table*}

\begin{figure*}[t]
  \centering
  \mbox{}
  \includegraphics[width=0.825\linewidth]{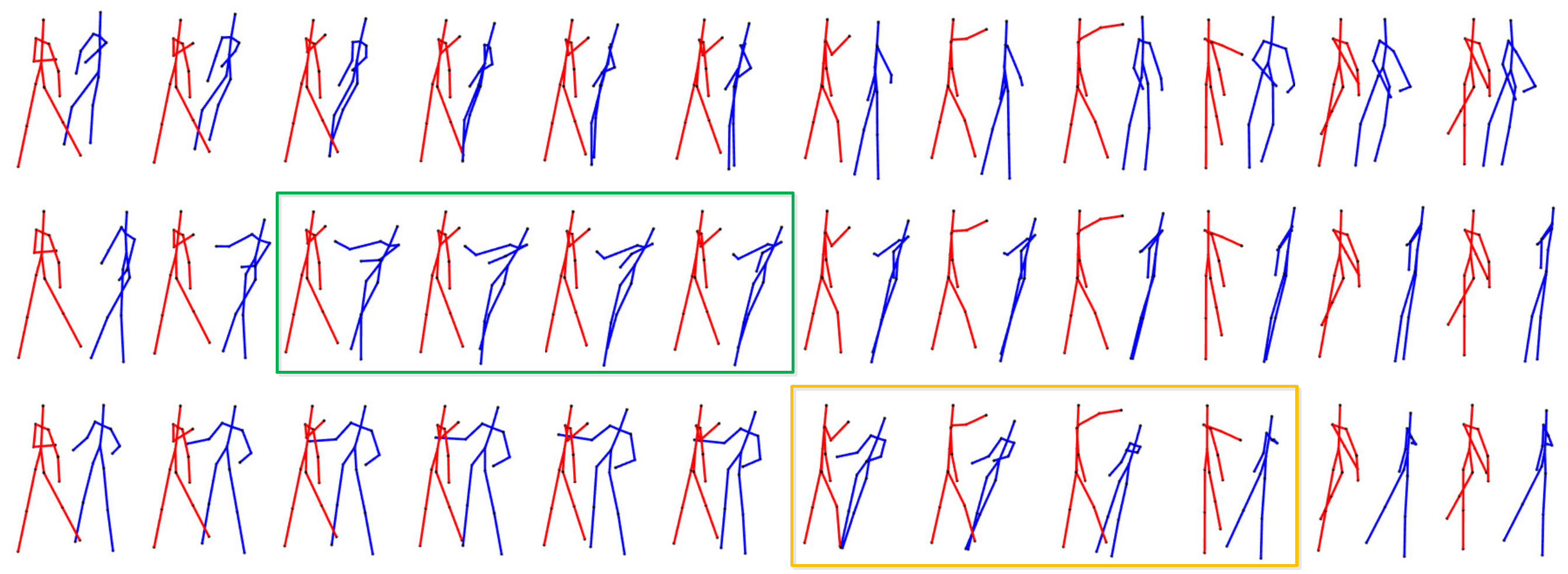}
  \includegraphics[width=0.8\linewidth]{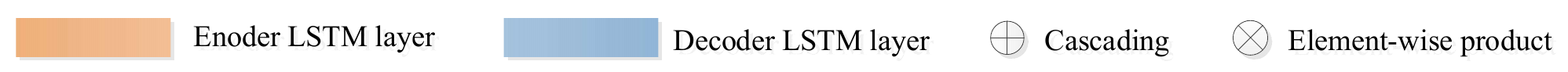}
\hfill \\
\hfill
  \includegraphics[width=.15\linewidth]{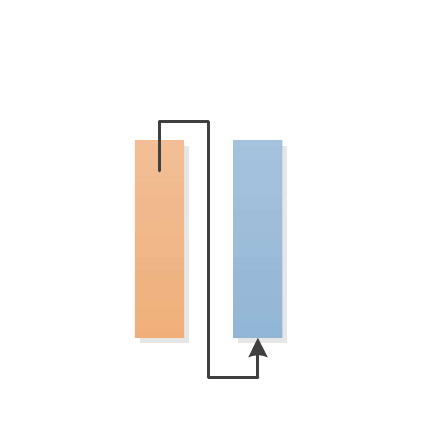}
\hfill 
  \includegraphics[width=.15\linewidth]{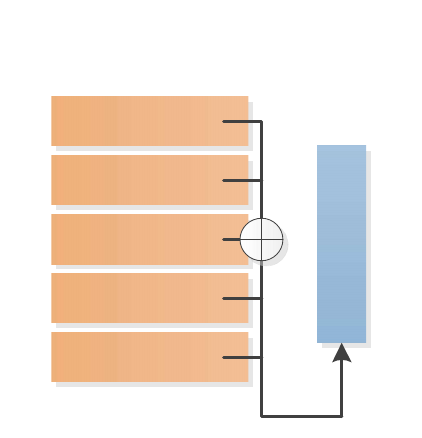}
\hfill 
  \includegraphics[width=.15\linewidth]{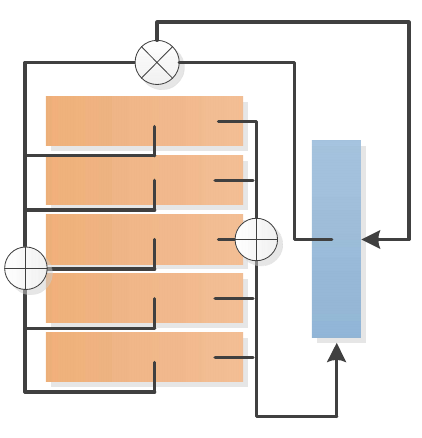}
\hfill\hfill\hfill \\
\qquad\qquad(a)\qquad\qquad\qquad\qquad\qquad\qquad\qquad(b)\qquad\qquad\qquad\qquad\qquad\qquad\qquad(c)\qquad\qquad
\mbox{}
  \caption{\label{fig:4} \textcolor{black}{Qualitative results and architectures of three generator modalities for the alignment test. The skeletons refer to the synthesized frames of a pushing reaction sequence in the SBU dataset. The top to the third rows are generated by methods (a) Seq2seq Generator, (b) Seq2seq Part-based Generator, and (c) Seq2seq Part-based Attentive Generator (our $G$). The green box highlights the biased frames, and the orange box highlights the aligned frames. We observe that when modeling the body part, the reactive motion shows less spatial artifacts, and further including the attentive mechanism can better align the two characters.}}
\end{figure*}

\begin{figure}[t]
\centering\includegraphics[width=1.\linewidth]{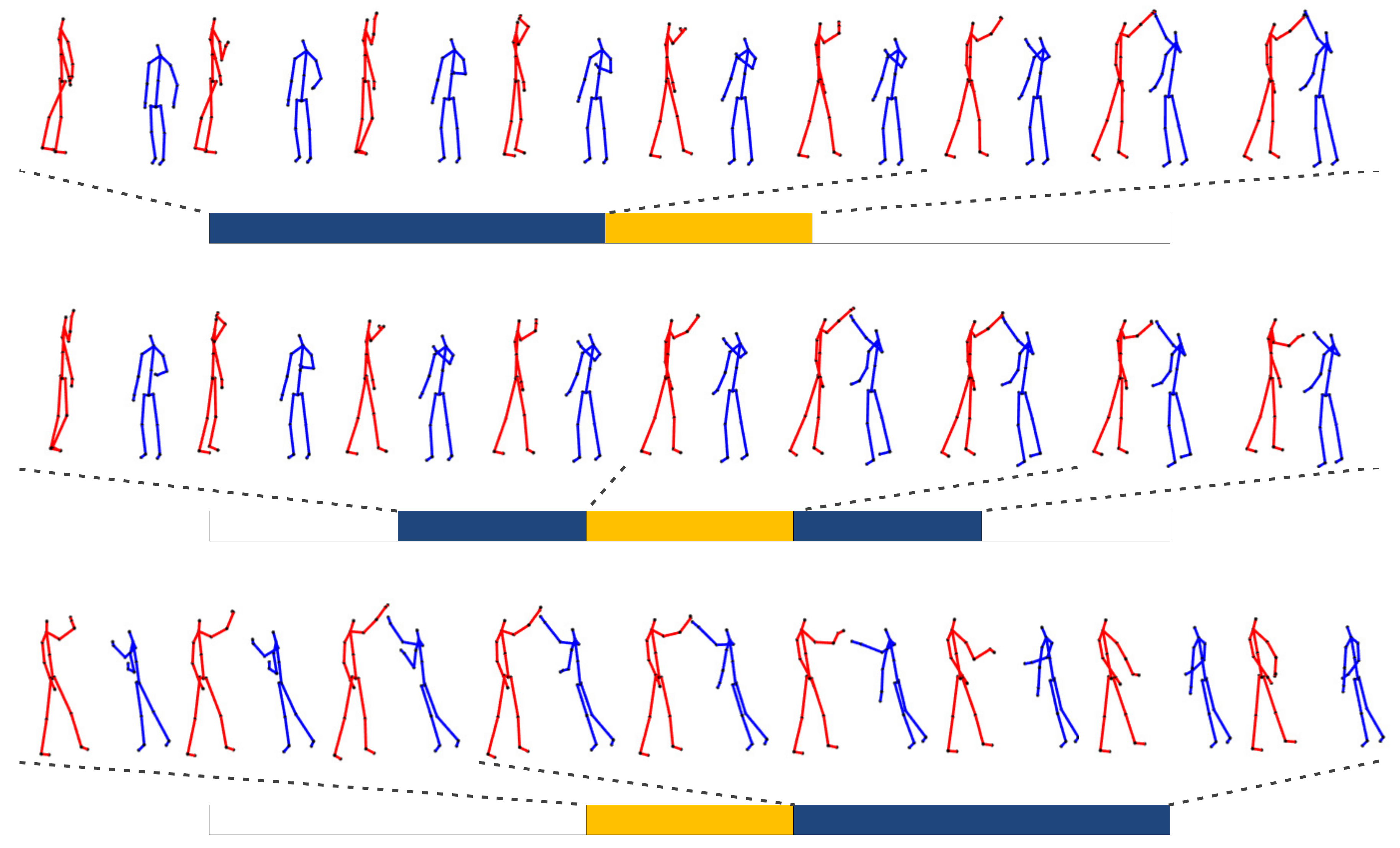}
\caption{\textcolor{black}{Interaction alignment demonstration of three phases from one high-five sequence in HHOI. The blue bar implies the individual time period and the orange bar is the overlap period which shows the keyframes of this interaction. For different time periods, the synthesized character aligns the input character with coincident arm movements.}}
\label{fig:6} 
\end{figure}

\begin{figure}
\centering\includegraphics[width=1.0\linewidth]{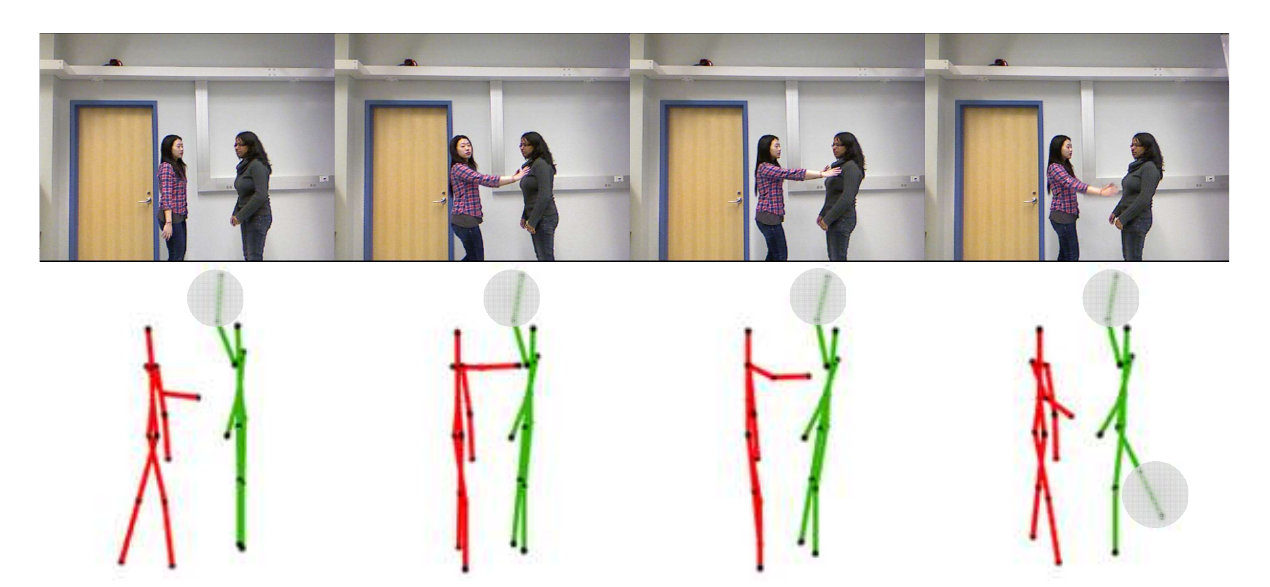}
\caption{Example skeleton errors in the SBU dataset. The grey area displays the inaccurate joint positions.}
\label{fig:7} 
\end{figure}

\subsection{Quantitative Evaluations}
\label{sec:4.3}
We also conduct quantitative analysis to test the effectiveness of the multi-class discriminator. The deterministic metric Average Frame Distance (AFD) is adopted to measure the geometric similarity between the learned skeleton $\hat{x}^{B}$ and the ground truth $x^{B}$, which is defined by: 
\begin{equation}
AFD\coloneqq \frac{1}{T}\sum_{t}\lVert\hat{x}_{t}^{B}-x_{t}^{B}\rVert^2.
\end{equation} 
The AFD comparison towards $D_{m}$ under different interaction class is shown in Table~\ref{tab:afd_dm}. We can see that the synthesized reactive motion shows a much lower positional error in most classes by including the multi-class classifier $D_m$, which verifies that discriminating different interactions helps improve the synthesized reactions with better quality.

\textcolor{black}{To evaluate our model, We also compare the proposed reactive synthesis method with prior work~\cite{huang2014action,huang2015approximate} that are closely related to ours, and some classic machine learning-based methods. Following~\cite{huang2015approximate}, the first baseline we adopted is the Nearest Neighbour~\cite{cover1967nearest} (denoted as NN) based on the framewise co-occurrence without considering temporal correlations. The second baseline is hidden Markov model~\cite{rabiner1986introduction} (denoted as HMM), which restores the reactive poses with sequential state transition based on the given movement. The third baseline is discrete Markov decision process~\cite{kitani2012activity} (denoted as DMDP) by discretizing the time steps with unsupervised clustering. In addition, we also compare with~\cite{huang2014action} and~\cite{huang2015approximate} that adopting kernel-based reinforcement learning (denoted as KRL) and maximum-entropy inverse optimal control (denoted as ME-IOC), respectively, for reaction synthesis.}

\textcolor{black}{The comparison results on different action classes are given in Table~\ref{tab:afd}. We observe that our method achieves comparable performance with the lowest prediction errors in half of the categories. For the interactions of \emph{pushing} and \emph{shaking hands}, the AFD differences between our method and the corresponding best models (i.e. DMDP and ME-IOC, respectively) are less than 0.1. Different from other actions, \emph{hugging} shows a relatively higher AFD with our model. This is because the large diversity caused by frequent self-occlusions makes it hard to learn the feature co-occurrence in this class, thus reducing the synthesis performance. Although the quantitative results are compatible with the statistical models~\cite{huang2014action} and~\cite{huang2015approximate}, their methods mainly sample or assemble source movements from the training data. This makes them less likely to be generalized to large-scale motions when more variations are needed in the synthesis to meet diverse user requirements.}

Furthermore, we quantify the recognition accuracy of the reaction generated by different combinations of losses. We first construct a two-layer LSTM with 512 units each layer and a linear layer connected to its end as the baseline classification network, and train it with the 3D joints of real interactions with the same cross-subject strategy as we train the reactive motion generator. The test interactions consist of real actions for character A and their corresponding real or synthesized reactions for character B. For this baseline evaluation, we denote it as \emph{prototype}. We also evaluate the model under different loss combinations: Adversarial loss only (denoted as $Adv.$), adversary with bone losses (denoted as $Adv.+\mathcal{L}_{skl}$), adversary with bone and continuity losses (denoted as $Adv.+\mathcal{L}_{skl}+\mathcal{L}_{con}$), and adversary with all 3 losses (denoted as $Adv.+\mathcal{L}_{skl}+\mathcal{L}_{con}+\mathcal{L}_{1}$).

The recognition performance on each interaction category of the two datasets is given by Table~\ref{tab:1} and \ref{tab:2}. In general, the discriminability will increase when we include more restrictions on the synthesized actor, and our model with all three constraints outperforms others, which shows the effectiveness and indispensability of each proposed loss function. For SBU dataset (Table~\ref{tab:1}), it is challenging to differentiate \emph{pushing} and \emph{punching} as the two reactions behave visually similar in skeletal representation, and it will mainly rely on the contractive $\mathcal{L}_{1}$ loss to examine the slight distinction in spatial patterns existed in two kinds of reactions. Another observation is that our architecture does not perform well in neutral types of interaction especially \emph{hug} since large biases of the bone lengths and frame jumping problems occurred because of abundant occlusions and intersects between two characters during hugging frames in the training set. This distortion makes the generator hard to learn its intrinsic spatial regularities and temporal dependencies. We also observed that the \emph{Shake hands} and \emph{Exchange object} interactions are highly similar and result in relatively low classification accuracy in those classes. Nevertheless, such ambiguity does not have a significant impact on the visual quality of the synthesized interactions as those two interactions are very similar in terms of body movements. In Table~\ref{tab:2}, the recognition results on HHOI with all types of losses are also the closest to the compared \textit{prototype} baseline. 

\begin{figure*}[t]
\centering
\subfigure[Shake hands (\#1)]{
\includegraphics[width=0.32\textwidth]{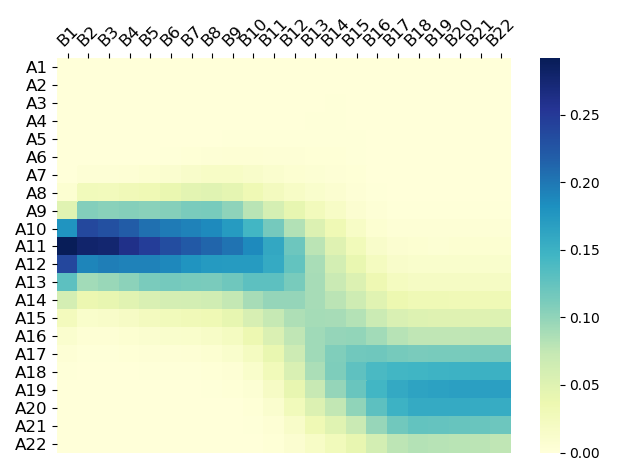}
}
\subfigure[Shake hands (\#2)]{
\includegraphics[width=0.32\textwidth]{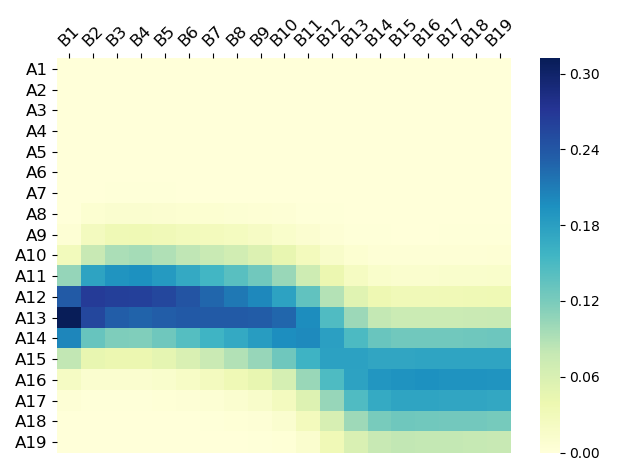}
}
\subfigure[Exchange objects]{
\includegraphics[width=0.32\textwidth]{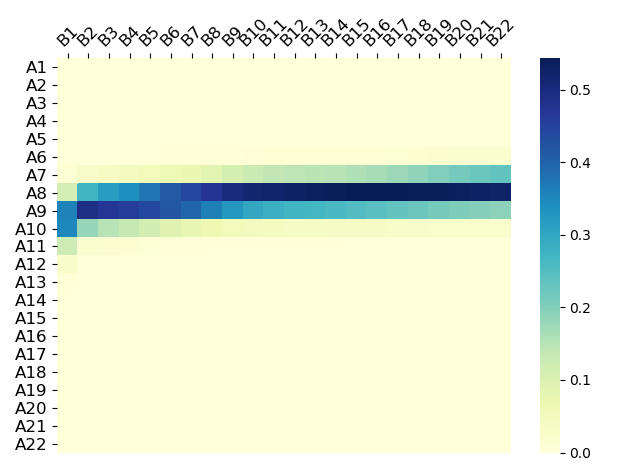}
}

\caption{\textcolor{black}{The example attention maps between the input character A and synthesized character B at every frame. (a) and (b) are attention maps of two \textit{shake hands} interactions, and (c) is \textit{exchange objects}, respectively. Note that the size of attention map may be varied based on the length of the interaction sequence.}}
\label{fig:attention_map}
\end{figure*}

\begin{figure*}
\centering
    \begin{minipage}{0.08\textwidth}
        (a)\quad\rotatebox[origin=c]{90}{Dodge \#1}
    \end{minipage}
    \begin{minipage}{0.9\textwidth}
        \includegraphics[width=1\linewidth]{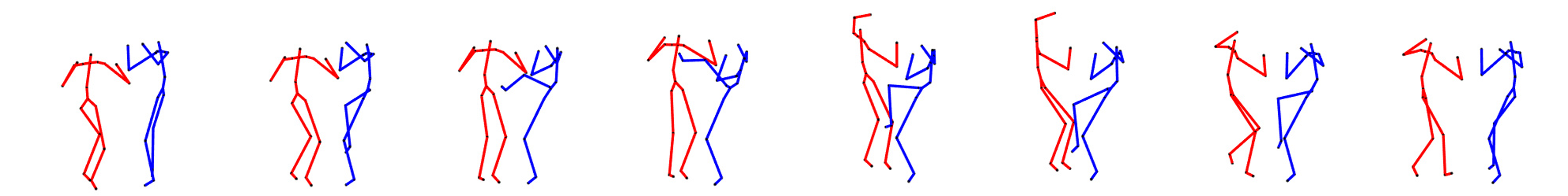}
    \end{minipage}
    \begin{minipage}{0.08\textwidth}
        (b)\quad\rotatebox[origin=c]{90}{Dodge \#2}
    \end{minipage}
    \begin{minipage}{0.9\textwidth}
        \includegraphics[width=1\linewidth]{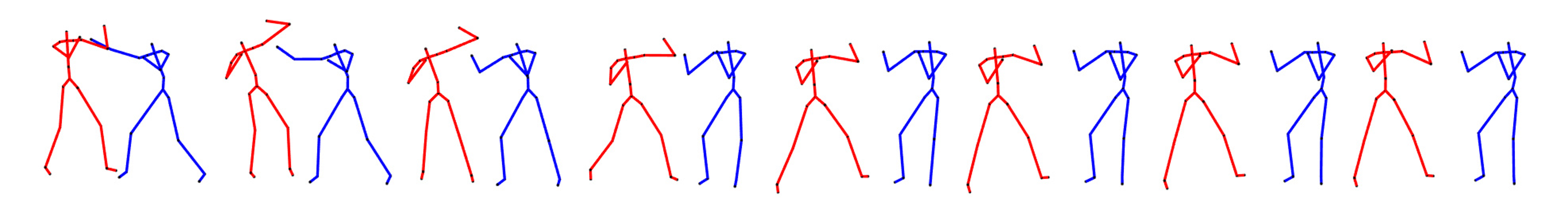}
    \end{minipage}
    \begin{minipage}{0.08\textwidth}
        (c)\quad\rotatebox[origin=c]{90}{High-five}
    \end{minipage}
    \begin{minipage}{0.9\textwidth}
        \includegraphics[width=1\linewidth]{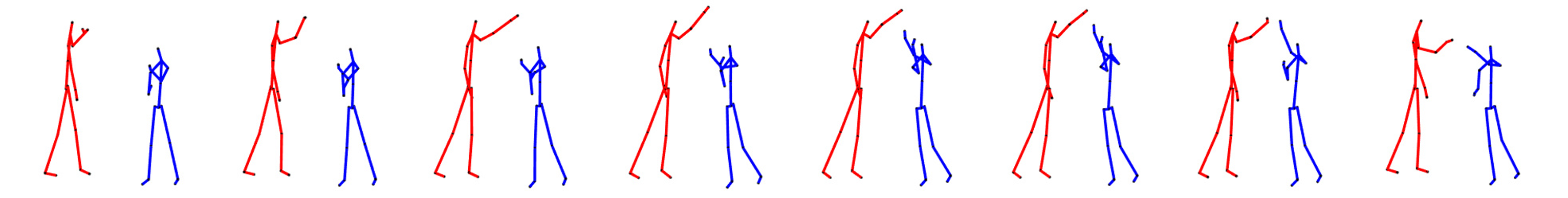}
    \end{minipage}\\
    \caption{\label{fig:more_test}\textcolor{black}{Example frames of the synthesized actions (blue character) by feeding in the reaction (red character).}}
\end{figure*}

\subsection{Interaction Alignment Evaluations}
\label{sec:4.2}
To clarify how each component of the generator structure contributes to the final output, we compare three ablation strategies on the network construction to train the generator $G$. The baseline structure (denoted as ``Seq2seq Generator") is formed by a two-layer LSTM with basic sequential encoder-decoder architecture. The second structure is trained with five LSTM layers separately, each of which encodes the action of an articulated branch in a skeleton, and their final states are cascaded to be interpreted by the decoder (denoted as ``Seq2seq Part-based Generator"). The third one is our method with the attention mechanism equipped with the encoder-decoder structure based on the second model (denoted as ``Seq2seq Part-based Attentive Generator"). 

The corresponding architecture and their visualized effects are compared in Fig.~\ref{fig:4}. We observe that when adding spatial hierarchy (the 1\textsuperscript{st} and 2\textsuperscript{nd} row), the encoder can better recognize the input action and react with less floating and artifacts. However, in the 2\textsuperscript{nd} row which temporal attention is not considered, we observe that the right character (synthesized) dodges before the left character (input) pushes. For the essential pushing frames, the right agent stops moving back and recovers gradually, which shows the misalignment in the whole interaction performance. As highlighted in the orange box of the 3\textsuperscript{rd} row, we can see that the temporal attention better aligns the movements of the two characters by dodging at a proper time, since the decoder can learn which interaction stage should the system pay more attention to for a punctual reaction.

We further test on three time phases of a high-five sequence as shown in Fig.~\ref{fig:6} (i.e. raise arm, high-five, put down arm). The synthesized reaction shows coincident arm raising and putting down with the input character in each time scope, which also demonstrates that our system can build the reaction based on the observed spatial pattern, but not answer back with a uniform temporal pattern. It indicates that the proposed network can not only identify and encode the detected context, but also provide real-time and refined feedback.

\textcolor{black}{To clarify the attention module, we also show the learned attention weights of three interaction samples from \textit{shake hands} and \textit{exchange objects}. As given in Fig.~\ref{fig:attention_map}, each element $\alpha(i,j)$ from Eq.~\ref{eq:addressing} in an attention map represents the attention value between character A in the $i$\textsuperscript{th} frame (i.e. A$i$) and character B in the $j$\textsuperscript{th} frame (i.e. B$j$). Since the attention is attached to the reaction, the active frames of A will contribute to the entire action of B. From Fig.~\ref{fig:attention_map}(a) and Fig.~\ref{fig:attention_map}(b), the wide range of non-zero weights indicates that the shaking interaction remains active for a long time, and it shows the alignment (higher values in diagonals) till the end of shaking. By comparing the two attention maps, we also observe that the attention pattern of different instances varies that is not determined by the interaction type. Compared to shaking hands, most of the large weights of exchanging objects are centered at a short period (i.e. A7$\sim$A10 in Fig.~\ref{fig:attention_map}(c)), which makes sense as the activity of exchanging is relatively fast.}

\textcolor{black}{Note that simply depending on the action type will generate some ambiguous reactive patterns (e.g. the 2\textsuperscript{nd} row in Fig.~\ref{fig:4}), while adding attention module helps to generate sample-wise reactive motion according to its received interaction pace. Thus, the advantage will also be kept even though the interaction shows less synchronization, such as \textit{waving back}.}

\subsection{\textcolor{black}{More Generalization Tests}}
\textcolor{black}{
We also conduct a generalization test by feeding in reactive motions in training and testing on unseen reactions. Some example generations are given in Fig.~\ref{fig:more_test}. By feeding in two dodging reactions (the red character), the model generates some attacking actions (the blue character), such as kicking and punching. When feeding in a high-five reaction, the model can recognize it and generate the high-five as well. We also observe that the system will not create some averaged action (e.g. kicking while punching) as the discriminator help to identify generation to a single type of response.}

\section{Limitations}
\label{sec:limitation}
For the limitations, the proposed model may fail to synthesize the microscopic movements when the interactions contain local actions. For example, during \emph{shaking hand} interaction, it is difficult to perform shaking for B's arm with the simple amplitude as A, which will result in a resemble acting as \emph{exchange object}. To reduce this ambiguity, the system is required to learn the geometric relationship between two actors to further reflect the reciprocal interaction in detail.

Another limitation of the method is that as a data-driven approach, the result of the synthesized motion will largely depend on the observed interaction in the dataset. \textcolor{black}{For example, feet floating may sometimes take the place of the walking steps in the generated kicking and dodging interactions. This is because, like many other deep learning-based action synthesis work~\cite{battan2021glocalnet,barsoum2018hp}, the walking pattern is hardly learned when most of the interactions observed are non-walking related. We improve the rendering using 3D stickman figures representing each bone with volumetric cylinders in the video, where the root positions are also included with less feet sliding. However, as an extension, it would be possible to fix this problem by constraining the velocity of the toe or heel when considering foot contact parameters in locomotion~\cite{holden2016deep}.} Due to the limitation of depth sensors, it is inevitable to draw in some occlusions and artifacts (e.g. Fig.~\ref{fig:7}), especially for the interactions with close contact such as $hugging$ which results in inaccuracies in the captured data. This will make it hard for the generated reaction to perform in the way of a true human motion. \textcolor{black}{Furthermore, the model proposed in this work uses 3D joint positions for motion synthesis. Because of the nature of the data, it is hard to fully synthesize a skinned character pose due to the impossibility to determine the orientation of the body joints. }

\section{Conclusion}
\label{sec:conclusion}
In this paper, we proposed an innovative human reaction generation system based on seq2seq generative adversarial network. The generator is self-adaptive which can autonomously recognize the observed action from spatial and temporal perspectives without the label information, and further shape a precise reaction. The dual discriminator with the binary and multi-class classifiers are designed to promote the authenticity and the discrimination of the reaction. The movements of body parts are analyzed hierarchically to discover the part-based features, and they are integrated to be interpreted by the decoder. An attention mechanism is also attached to the decoder to align the synthesized interaction. To synthesize a more realistic reaction, we add a skeleton loss to keep the basics of the physical body structure, a continuity loss to smooth the appearance among motion frames and a contractive loss to reduce the artifacts of the generated movements. 

We have both qualitatively and quantitatively evaluated our reaction synthesis approach with respect to the discriminability, the synchronism between characters, and the similarity to the actual reaction. Experimental results show that the proposed generative model can produce logically and numerically analogous generations of human reaction when the input action is provided.


\textcolor{black}{In this work, we synthesize the natural reactive patterns by assuming the action and reaction appear in pairs. Since human responses in social interaction should not be limited to one single reaction pattern, as future work, we aim to increase the diversity of the generated reactive motion. Possible solutions include disentangling the basic reactive patterns and different reactive styles, or accommodating random noise $z$ to our generative model to increase the variations of the synthesized reaction.} In addition, creating an online human reactive motion with local temporal attention is another interesting direction to explore.

\textcolor{black}{As another potential future direction, our work can be further improved by collecting a larger interaction dataset where the distribution-based metrics such as FID (Fréchet Inception Distance)~\cite{heusel2017gans} can be applied to evaluate the generation space.}

\bibliographystyle{cag-num-names}
\bibliography{refs}


\end{document}